\documentclass[superscriptaddress,aps,pra,twocolumn,nofootinbib,babel]{revtex4-1}
\usepackage{graphicx}
\usepackage{amsmath}
\usepackage{amssymb}
\usepackage{comment}
\usepackage{placeins}
\usepackage[caption=false]{subfig}
\usepackage[colorlinks]{hyperref}
\usepackage{hypcap}
\usepackage[english]{babel}
\usepackage{xcolor} 
\usepackage{gensymb}

\newcommand{\eq}[1]{Eq.~\hyperref[eq:#1]{(\ref*{eq:#1})}}
\renewcommand{\sec}[1]{\hyperref[sec:#1]{Section~\ref*{sec:#1}}}
\newcommand{\app}[1]{\hyperref[app:#1]{Appendix~\ref*{app:#1}}}
\newcommand{\tab}[1]{\hyperref[tab:#1]{Table~\ref*{tab:#1}}}
\newcommand{\fig}[1]{\hyperref[fig:#1]{Figure~\ref*{fig:#1}}}
\newcommand{\figa}[2]{\hyperref[fig:#1]{Figure~\ref*{fig:#1}#2}}
\newcommand{\figx}[2]{\hyperref[fig:#1]{Figure~\ref*{fig:#1}(#2)}}
\newcommand{\thm}[1]{\hyperref[thm:#1]{Theorem~\ref*{thm:#1}}}
\newcommand{\lem}[1]{\hyperref[lem:#1]{Lemma~\ref*{lem:#1}}}
\newcommand{\cor}[1]{\hyperref[cor:#1]{Corollary~\ref*{cor:#1}}}
\newcommand{\defn}[1]{\hyperref[def:#1]{Definition~\ref*{def:#1}}}
\newcommand{\alg}[1]{\hyperref[alg:#1]{Algorithm~\ref*{alg:#1}}}
\newcommand{\specialcell}[2][c]{%
  \begin{tabular}[#1]{@{}c@{}}#2\end{tabular}}

\def\bra#1{\mathinner{\langle{#1}|}}
\def\ket#1{\mathinner{|{#1}\rangle}}
\newcommand{\braket}[2]{\langle #1|#2\rangle}

\newcommand{\ignore}[1]{}

\newcommand{\be}{\begin{equation}}
\newcommand{\ee}{\end{equation}}
\newcommand{\ba}{\begin{eqnarray}}
\newcommand{\ea}{\end{eqnarray}}
\usepackage{times}

\newcommand{\jrf}[1]{{\color{black}{#1}}}

\input{Qcircuit}

\begin{document}

\title{Strategies for quantum computing molecular energies using the unitary coupled cluster ansatz}

\author{Jonathan Romero}
\affiliation{Department of Chemistry and Chemical Biology, Harvard University, Cambridge, MA 02138}

\author{Ryan Babbush} 
\affiliation{Google Inc., 340 Main Street, Venice CA 90291}

\author{Jarrod R. McClean} 
\affiliation{Google Inc., 340 Main Street, Venice CA 90291}

\author{Cornelius Hempel}
\affiliation{ARC Center for Engineered Quantum Systems, School of Physics, The University of Sydney, NSW 2006 Australia}

\author{Peter Love}
\affiliation{Department of Physics and Astronomy, Tufts University, Medford, MA 02155}

\author{Al\'an Aspuru-Guzik}
\email[Corresponding author: ]{aspuru@chemistry.harvard.edu}
\affiliation{Department of Chemistry and Chemical Biology, Harvard University, Cambridge, MA 02138}

\date{\today}

\begin{abstract}
The variational quantum eigensolver (VQE) algorithm combines the ability of quantum computers to efficiently compute expectation values with a classical optimization routine in order to approximate ground state energies of quantum systems. In this paper, we study the application of VQE to the simulation of molecular energies using the unitary coupled cluster (UCC) ansatz. We introduce new strategies to reduce the circuit depth for the implementation of UCC and improve the optimization of the wavefunction based on efficient classical approximations of the cluster amplitudes. Additionally, we propose an analytical method to compute the energy gradient that reduces the sampling cost for gradient estimation by several orders of magnitude compared to numerical gradients. We illustrate our methodology with numerical simulations for a system of four hydrogen atoms that exhibit strong correlation and show that the circuit depth of VQE using a UCC ansatz can be reduced without introducing significant loss of accuracy in the final wavefunctions and energies. 
\end{abstract}

\maketitle

\section*{Introduction}

The solution to the time-independent Schr\"odinger equation for molecular systems allows for the prediction of chemical properties, holding the key to materials discovery and catalyst design \cite{Curtarolo.NM.12.191.2013,huskinson.N.505.7482.2014,Su.CS.6.885.2015,hachmann.JPCL.2.2241.2011}. Despite advances in the field of quantum chemistry, many relevant problems such as the prediction of chemical rates and the description of transition-metal complexes remain challenging \cite{Bell.MP.102.319.2004,marti.PCCP.13.6750.2011}. These difficulties stem from the approximate nature of classically tractable quantum chemistry approaches, which often fail in the description of strongly correlated systems \cite{Lyakh.CR.112.182.2011,Szalay.CR.112.108.2011}. In addition, the application of exact methods, such as exact diagonalization of the electronic Hamiltonian, require exponential resources with current classical algorithms, limiting the exact simulation of molecular energies to systems comprising only a few atoms \cite{Head-Gordon.PT.61.58.2008a,Helgaker2013}.

Feynman envisioned that quantum computers could provide a tractable way to simulate quantum systems \cite{Feynman.IJTP.21.467.1982}. This idea, formalized by Abrams and Lloyd a decade later \cite{Abrams.PRL.79.2586.1997}, has been developed into a series of quantum algorithms for quantum simulation \cite{Georgescu.RMP.86.153.2014,Kassal.ARPC.62.185.2011,Yung.2014.Chapter}. The first algorithm extending these approaches to the calculation of molecular energies was proposed by Aspuru-Guzik et al.~\cite{Aspuru-Guzik.S.309.1704.2005}. This first proposal, further developed in \cite{Whitfield.MP.109.735.2011}, combines Trotterization of the molecular Hamiltonian and phase estimation (PEA) to compute the ground state energy of a molecule. 

Early studies on the quantum resources required by this algorithm showed that the circuit depth scales as $O(N^8)$ \cite{Hastings.QIC.15.1.2015}, where $N$ the total number of spin-orbital functions. Fortunately, numerical studies indicated that the scaling for real molecules is closer to $O(N^6)$ \cite{Poulin.QIC.15.361.2015} or $O(Z_{\mathrm{max}}^3 N^4)$ when trying to simulate ground states. Here, $Z_{\mathrm{max}}$ is the largest nuclear charge of the molecule \cite{Babbush.PRA.91.22311.2015}. Recent proposals have developed new algorithms for this problem by considering simulation based on Taylor series methods as opposed to Trotterization \cite{BabbushSparse1,Kivlichan2017BoundingSpace}, performing simulations in a fixed particle number manifold \cite{BabbushSparse2,Toloui2013,BabbushSymmetry,Bravyi2017,Steudtner2017}, and considering specialized basis functions \cite{BabbushLow,Kivlichan2017}. Despite these recent theoretical improvements, all phase estimation based algorithms for this problem are unlikely to solve classically intractable molecules without error-correction. The variational quantum eigensolver (VQE)~\cite{Peruzzo.NC.5.4213.2014,Mcclean.NJP.18.023023.2016,Wecker.PRA.92.042303.2015} is a an alternative algorithm that is closer to near-team applicability due to lower coherence time requirements.

The VQE algorithm finds the best variational approximation to the ground state of a given Hamiltonian for a particular choice of ansatz. This task is achieved by two subroutines. The first subroutine employs a quantum computer to prepare a parameterized wavefunction ansatz and measure the expectation value of the Hamiltonian given a set of values for the parameters. The second subroutine consists of an optimization algorithm running on a classical computer. The optimization algorithm employs the quantum subroutine as an objective function and finds the parameters that minimize the energy of the ansatz. This procedure offers several advantages that make it a candidate for exploiting the performance of near-future quantum devices:  adaptability to different quantum architectures, intrinsic robustness to quantum errors \cite{mcclean2016hybrid,Omalley.PRX.6.031007.2016} and a smaller coherence time requirements \cite{Mcclean.NJP.18.023023.2016}.

The VQE approach was first applied to the simulation of molecular energies. In this case, a trial wavefunction is prepared by the application of a parametrized unitary, followed by the calculation of the energy via Hamiltonian averaging \cite{McClean.JPCL.5.4368.2014,Mcclean.NJP.18.023023.2016}. The value of the energy is minimized using a classical optimization routine that updates the variational parameters. Accordingly, the final cost of the calculation depends on the number of iterations required for convergence and the amount of operations involved in each preparation and measurement cycle of the quantum subroutine. This optimization scheme has been experimentally demonstrated in different quantum platforms, including photonic chips \cite{Peruzzo.NC.5.4213.2014}, ion traps \cite{Shen.apa...2015,Hempel.inprep} and superconducting circuits \cite{Omalley.PRX.6.031007.2016,Kandala2017}.

Traditionally, a unitary coupled cluster (UCC) approach has been used as the ansatz for the state preparation \cite{Peruzzo.NC.5.4213.2014,Yung.SR.4.3589.2014,Mcclean.NJP.18.023023.2016}. This method provides a hierarchy of wavefunctions that can be prepared on a quantum computer using a polynomial number of gates and it is believed to provide better accuracy than classical coupled cluster \cite{Kutzelnigg.1977.Chapter,Hoffmann.JCP.88.993.1988,Bartlett.CPL.155.133.1989,Cooper.JCP.133.234102.2010,Evangelista.JCP.134.224102.2011}, which is generally regarded as the ``gold standard" of quantum chemistry \cite{Bartlett.RMP.79.291.2007}. Despite these advantages, recent studies have pointed out that the number of parameters in UCC might be still too large to allow practical calculations for large molecules \cite{Wecker.PRA.92.042303.2015}. 

In this paper, we aim to describe in more detail the implementation of VQE approaches for molecular systems using a UCC ansatz and introduce strategies to improve its efficiency. In \sec{TheoryBackground}, we describe the approaches commonly used in classical quantum chemistry calculations and introduce the UCC ansatz in this context. In \sec{vqeucc}, we discuss in detail the implementation of VQE with a UCC ansatz, including the generation of initial guesses and the reduction of computational resources using pre-screening of the cluster amplitudes and active space approaches. In addition, we introduce a method to compute the gradient of the energy with respect to the variational parameters that can be combined with gradient-based optimization methods. In \sec{numerics}, we illustrate the proposed strategies through numerical simulations of the VQE approach for a variety of chemical systems. Finally, in \sec{discussion} we present a brief discussion of the results.

\section{Background}\label{sec:TheoryBackground}

\subsection{Quantum chemistry in second quantization}

Within the Born-Oppenheimer approximation, a molecule is comprised of a system of $\eta$ electrons interacting in the potential produced by nuclei located at fixed positions. We may describe this problem using the formalism of second quantization, where $N$ single-particle spin orbitals can be either empty or occupied. 
Any interaction between electrons can be represented using annihilation and creation operations, $a_p$ and $a^{\dagger}_p$, that obey the following anti-commutation relations, associated with fermionic statistics:
\begin{align}\label{eq:anticommutation}
[a_j,a_k]_{+}=0 \quad [a_j^{\dagger},a_k^{\dagger}]_{+}=0 \quad [a_j,a_k^{\dagger}]_{+}=\delta_{jk}
\end{align}
where $[a,b]_{+} \equiv ab+ba$. In the absence of external fields the non-relativistic molecular Hamiltonian can be written as:
\begin{align}\label{eq:molHamiltonian}
H=h_{nuc}+\sum_{pq}h_{pq} a^{\dagger}_p a_q + \frac{1}{2} \sum_{pqrs} h_{pqrs} a^{\dagger}_p a^{\dagger}_q a_r a_s
\end{align}
where $h_{nuc}$ corresponds to the classical electrostatic repulsion between nuclei, and the constants $h_{pq}$ and $h_{pqrs}$ correspond to the one- and two-electron integrals. Using atomic units, where the electron mass $m_e$, the electron charge $e$, Bohr radius $a_0$, Coulomb's constant and $\hbar$ are unity, we may write:
\begin{align}
&h_{pq} = \int d\sigma \varphi_p^*(\sigma) \left(-\frac{\nabla_{\vec{r}}^2}{2} - \sum_{i} \frac{Z_i}{|\vec{R}_i - \vec{r}|} \right)\varphi_q(\sigma) \label{eq:single_int}\\
 &h_{pqrs} = \int d\sigma_1\ d\sigma_2\ \frac{ \varphi_p^*(\sigma_1) \varphi_q^*(\sigma_2)  \varphi_s(\sigma_1) \varphi_r(\sigma_2) }{|\vec{r}_1 - \vec{r}_2|} \label{eq:double_int}\\
&h_{nuc}=\frac{1}{2}\sum_{i \neq j} \frac{Z_i Z_j}{|\vec{R}_i-\vec{R}_j|}
\end{align}
Here $Z_i$ represents the nuclear charge, $\vec{r}$ and $\vec{R}$ denote electronic and nuclear spatial coordinates, respectively, and $\sigma$ is now a spatial and spin coordinate with $\sigma_i=(\vec{r}_i;s_i)$. Summations run over all nuclei. The function $\varphi(\sigma)$ represent one-electron functions (spin-orbitals) that are often obtained from a mean field calculation such as Hartree-Fock (HF).

After removing the translational and rotational degrees of freedom, the electronic energy of a molecular system is a function of $3q-6$ parameters ($3q-5$ for linear molecules) that we will denote by $\vec{\bf R}$, where $q$ is the number of atoms. The function $E(\vec{\bf R})$ is called the potential energy surface (PES). The accurate calculation of the PES is one of the main challenges of quantum chemistry as it is required for predicting and understanding a wide range of chemical processes, such as reaction dynamics, bond-breaking and chemical kinetics. 

The prediction of thermochemical properties such as reaction rates determines the accuracy required from ab initio calculations of the PES \cite{peterson.TCA.131.1.2012}. Chemical rates, for instance, are exponentially sensitive to changes in the Gibbs free energy, and thus changes in the PES. This sensitivity can be seen from the Erying equation for chemical rates,
\begin{equation}
\textrm{rate} \propto \frac{ e^{-\beta \Delta G^{\ddagger}}}{\beta},
\end{equation}
where $\Delta G^{\ddagger}$ is the difference in free energy between reactants and transition state and $\beta$ is the inverse temperature in atomic units. At room temperature and atmospheric pressure, an error $\epsilon$ in $ \Delta G^{\ddagger}$ of 1.4 kcal/mol translates to a chemical rate error of a factor of ten. This leads to the definition of {\sl chemical accuracy} which sets $\epsilon$ to the order of 1 kcal/mol or approximately $1.59\times 10^{-3}$ Hartrees ($43.3$ meV) \cite{Helgaker2013}.

\subsection{Classical ab initio approaches to quantum chemistry}

The inherent difficulty of solving the Schrodinger equation for many-electron systems has motivated the development of a series of standard models for the construction and calculation of approximate electronic wavefunctions in quantum chemistry. The simplest approach is to represent the wavefunction as a single anti-symmetrized product of one-electron functions, known as a Slater determinant. The Hartree-Fock method provides such a single-determinant solution. In this scheme, the molecular orbitals are expressed as a linear combination of atomic orbital functions. The combination coefficients are then optimized by a self-consistent variational procedure in which each particle is made to interact with the average density of the other particles. The output of this calculation provides a mean-field approximation to the molecular wavefunction. Unfortunately, the Hartree-Fock method is incapable of approximating the electron correlation effects that are essential for computing energies within or close to chemical accuracy \cite{Helgaker2013}.

To correct for this problem, one can expand the wavefunction as a superposition of all the determinants in the $\eta$-electron Fock space. The coefficients in the expansion can be parametrized in different ways, defining different models for the description of electron correlation. Two popular parametrizations are the configuration interaction (CI) and the coupled-cluster (CC) methods. 

In the full configuration interaction (FCI) approach, which is exact within a given basis, the wavefunction is expanded as a linear combination of all the determinants in the $\eta$-Fock space. The coefficients of the expansion can be solved for by variational minimization of the energy, providing the exact wavefunction for a given orbital basis. Unfortunately, the FCI wavefunction becomes rapidly intractable due to the factorial dependence on the number of determinants $N$ related to the total number of spin orbitals \cite{Helgaker2013}.

To generate classically-tractable CI approaches one can truncate the CI expansion to include only determinants with a fixed number of excitations with respect to a reference configuration. The reference is usually chosen to be the Hartree-Fock state. This idea can be formalized by defining excitation operators as follows:

\begin{align}\label{eq:Tsum}
T&=\sum^{\eta}_{i=1} T_i\\
T_1&=\sum_{\substack{i \in \text{occ}\\ a \in \text{virt}}} t^{i}_{a} a^{\dagger}_{a} a_{i} \\
T_2&=\sum_{\substack{i>j\in \text{occ}\\ a>b \in \text{virt}}} t^{i j}_{a b} a^{\dagger}_{a} a^{\dagger}_{b} a_{i} a_{j}\\
\nonumber&\ldots 
\end{align}
where the $occ$ and $virt$ spaces are defined as the occupied and unoccupied sites in the reference state. In this construction, the operator $T_1$ generates single excitations from the reference, $T_2$ generates double excitations and the definition of higher order excitations follows naturally. $t^{i}_{a}$ and $t^{i j}_{a b}$ correspond to expansion coefficients. The exact full CI wavefunction is thus,
\begin{align}
\ket{\textrm{FCI}} & = (1+T) \ket{\textrm{HF}}\\
E_{\textrm{FCI}} & = \min_{\vec{t}} \frac{\bra{\textrm{FCI}} H \ket{\textrm{FCI}}}{\braket{\textrm{FCI}}{\textrm{FCI}}}\nonumber
\end{align}
where $\ket{\textrm{HF}}$ is the reference state (for instance, the Hartree-Fock solution) and $\vec{t}$ is the vector comprising the expansion coefficients. The maximum number of excitations allowed, defines the order of truncation, $k$. The FCI solution can be systematically approached by increasing $k$. The computational cost of truncated single-reference CI approaches scales as $O(\eta^{k} (N-\eta)^{k+2})$, assuming $N,\eta>>k$. Tractable classical CI truncation is generally limited to single and double excitation operators, which define the CI singles and doubles method (CISD). 

The truncated CI expansion suffers from two major problems. First, the method converges slowly when applied to highly correlated systems. To circumvent this problem we can use an entangled reference state that captures the main computational states contributing to the total wavefunction. This is the base of  \emph{multireference}\ methods in quantum chemistry \cite{Helgaker2013,Szalay.CR.112.108.2011}, 
which are generally more involved than truncated single reference CI approaches.

The second complication is that configuration interaction is not size-extensive. A method that is size-extensive for a system of non-interacting fragments has a wavefunction that is multiplicatively seperable and an energy that is proportional to the size of the system \cite{Helgaker2013}. This means that the total wavefunction factorizes as a product of the wavefunctions of the independent fragments and the corresponding energy is the sum of the energies of the fragments. These conditions assure that the energy scales linearly with the size of the system. Size-extensivity is a desirable feature for approximate methods in quantum chemistry because many chemical properties, such as the atomization energy, are obtained by subtracting the energy of systems with different sizes. In addition, we expect that higher order expansions must be used for larger molecules if the method is not size-extensive.

The lack of size-extensivity of the truncated CI wavefunction can be overcome by recasting the linear FCI parametrization in the form of a product wavefunction. This is done in the CC method by means of an exponential ansatz:
\begin{align}
\ket{\Psi} = e^{T} \ket{\textrm{HF}}
\end{align}
where the operator $T$ is defined as for CI. Notice that in this scheme the parameters $\vec{t}$ constitute excitation amplitudes instead of expansion coefficients. As with CI, CC is usually truncated at some fixed level of excitation. For instance, the method known as coupled cluster singles and doubles (CCSD) is based on the ansatz,
\begin{equation}
\ket{\textrm{CCSD}} = e^{T_1 + T_2} \ket{\textrm{HF}}.
\end{equation}
Whereas truncated CI wavefunctions contain contributions from a polynomial number of determinants at a given truncation level, truncated CC wavefunctions have support on all the determinants in the $\eta$-Fock space. Tractable implementations of the coupled-cluster theory rely on projecting the Schr\"{o}dinger equation in the form
\begin{equation}
e^{-T} H e^{T} \ket{\textrm{HF}} = E_\mathrm{CC} \ket{\textrm{HF}}
\end{equation}
against a set of configurations $\{\bra{\mu}\}$. This set spans the space of all the states that can be reached by applying the truncated cluster operator $T$ linearly to the reference state \cite{Bartlett.RMP.79.291.2007}.  This treatment generates the following set of non-linear equations for the CC energy and amplitudes:
\begin{align}\label{eq:ccprojectede}
\bra{\textrm{HF}} e^{-T} H e^{T} \ket{\textrm{HF}} = E \\
\bra{\mu} e^{-T} H e^{T} \ket{\textrm{HF}} = 0
\end{align}
The key point in establishing the size-extensivity of CC theory is to note that the operator $e^{-T} H e^{T}$, known as the similarity-transformed Hamiltonian, is additively separable and produces additively separable energies. Similarly, it can be shown that the operator $e^{T}$ is multiplicatively separable and thus generates multiplicatively separable wavefunctions \cite{Helgaker2013}.

In practice, the similarity-transformed Hamiltonian is expanded using the  Baker-Campbell-Hausdorff (BCH) formula:
\begin{align}\label{eq:BCHCCC}
e^{-T} H e^{T} = & H + \left[H, T\right] + \frac{1}{2}\left[\left[H, T\right], T\right] \notag \\
& + \frac{1}{3!} \left[\left[\left[H, T\right] T\right], T\right] + \frac{1}{4!} \left[\left[\left[\left[H, T\right] T\right], T\right], T\right].
\end{align}
The expansion terminates at fourth order due to the commutation properties of excitation operators for the special case that the reference is a single determinant \cite{Bartlett.RMP.79.291.2007,Helgaker2013}. This fact allows for an efficient evaluation of the projected CC equations without further approximation.

While truncated CC is classically tractable and more accurate than truncated CI, there are two substantial weaknesses to the theory. The first weakness is the BCH expansion of the similarity-transformed Hamiltonian is only convergent under the assumption of a single reference state. Consequently, single reference coupled cluster generally performs poorly for strongly correlated systems. This means that coupled cluster is fairly reliable when computing energies at equilibrium configurations but likely to fail for transition states or near dissociation limits of multiple bonds. At those geometries, excited surfaces may become nearly degenerate with the ground state and a single determinant (e.g. the Hartree-Fock state) may have very small overlap with the ground state. Although the field of multireference coupled cluster methods has expanded in the last years, current approaches are still far from being practical for large molecular systems \cite{Lyakh.CR.112.182.2011}.

The second weakness of the projected coupled-cluster formulation is that the operator $e^{T}$ is not unitary and therefore the energy obtained from \eq{ccprojectede} is not variational. In the next section we discuss a formulation of coupled cluster theory that is variational and can be made multireference. While this formulation is not classically tractable, it can be implemented using a quantum computer.

\subsection{Unitary coupled cluster}

The shortcomings of the traditional coupled cluster ansatz described in the previous section can be overcome by redefining the excitation operator to be unitary, an approach known as unitary coupled cluster (UCC) \cite{Kutzelnigg.1977.Chapter,Hoffmann.JCP.88.993.1988,Bartlett.CPL.155.133.1989}:
\begin{equation}
\label{eq:ucc}
\ket{\Psi} = e^{T - T^{\dagger}} \ket{\textrm{HF}}.
\end{equation}
the total energy of the system is obtained from the variational principle as:
\begin{align}
\label{eq:UCCenergy}
E=\min_{\vec{t}}  \bra{\textrm{HF}} e^{-(T-T^{\dagger})} H e^{T-T^{\dagger}} \ket{\textrm{HF}}
\end{align}
while this ansatz is variational and spans the same Hilbert space as the original coupled cluster ansatz, \eq{ucc} does not lead to equations which can be tractably solved on a classical computer \cite{kutzelnigg.TCA.80.349.1991,taube.IJQC.106.3393.2006}. To see this we can examine the BCH expansion of the similarity transform hamiltonian for UCC:
\begin{align}\label{eq:BCHUCC}
 e^{T^{\dagger}-T} H  e^{T - T^{\dagger}} = & H + \left[H, T\right] + \left[T^{\dagger},H \right] + \frac{1}{2}(\left[\left[H, T\right], T\right] \notag \\
& + \left[T^{\dagger}, \left[T^{\dagger}, H\right]\right] + \left[H, \left[T, T^{\dagger}\right] \right]) + \cdots
\end{align}
In contrast with the expansion for CC (\eq{BCHCCC}), \eq{BCHUCC} involves terms that depend on the commutators between $T$ and $T^{\dagger}$ operators, for which there is no natural termination point \cite{kutzelnigg.TCA.80.349.1991,taube.IJQC.106.3393.2006}. Therefore, the BCH series for UCC is infinite and thus there is currently no known method for efficiently evaluating the energy and amplitude equations on a classical computer without further approximation.

Nonetheless, the minimization of the UCC ansatz is of great interest to the quantum chemistry community that has been trying to develop tractable approximations to this theory for many years \cite{Kutzelnigg.1977.Chapter,Hoffmann.JCP.88.993.1988,Bartlett.CPL.155.133.1989,Cooper.JCP.133.234102.2010,Evangelista.JCP.134.224102.2011}. Fortunately, the operator $e^{T - T^{\dagger}}$ can be readily applied on a quantum computer, which makes it possible to prepare UCC wavefunctions with truncated cluster expansions, as shown in \cite{Yung.SR.4.3589.2014,Peruzzo.NC.5.4213.2014,Mcclean.NJP.18.023023.2016}.

\section{Variational quantum eigensolver for UCC}\label{sec:vqeucc}

\begin{figure}
\centering
\includegraphics[width=9cm]{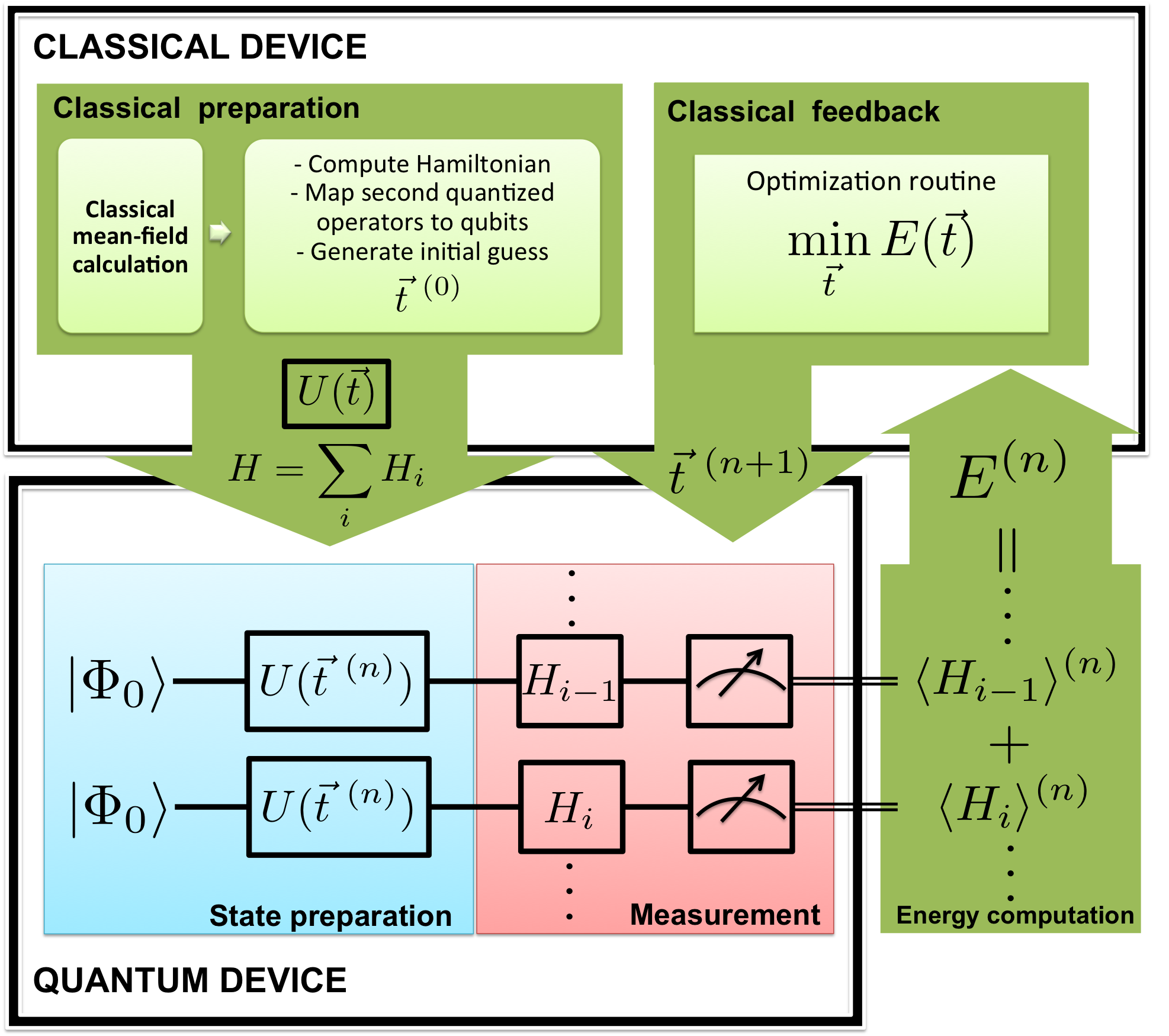}
%
\caption{Schematic representation of the Variational Quantum Eigensolver algorithm applied to the UCC ansatz. The classical optimization routine adds the expectation values of the Hamiltonian terms to calculate the energy and estimates a new value for the coupled cluster amplitudes, $\vec{t}$. The process is repeated until achieving  convergence on the total energy and $\vec{t}$.}\label{fig:UCCdiagram}
\end{figure}

The VQE algorithm comprises three iterative steps: 1) preparation of the wavefunction by application of parameterized state preparation unitaries; 2) determination of the expectation value of every term in the Hamiltonian via an efficient partial tomography \cite{McClean.JPCL.5.4368.2014} and 3) calculation of the total energy and determination of a new set of state preparation parameters in a classical computer. This scheme avoids the substantial overhead of quantum phase estimation that causes other quantum algorithms for chemistry to require very long coherent evolution. It also offers flexibility in the length of the circuit for state preparation, that depends on the choice of ansatz for the state preparation.

In the specific case of UCC, the preparation of the wavefunction encompasses two steps: preparation of the reference state, $|\Phi_{0}\rangle$, and application of the UCC unitary, $U(\vec{t})$, that prepares the UCC wavefunction. The algorithm starts with a guess of the UCC amplitudes, $\vec{t}^{\ (0)}$, and iteratively converges to a final set of parameter by variationally minimizing the energy. At the $n$-th iteration, the UCC wavefunction is prepared using $\vec{t}^{\ (n)}$ and the expectation value of the Hamiltonian, $H$, is obtained as the sum of the expectation values of all the terms, $\langle H \rangle = \sum_i \langle H_i \rangle $. The classical optimization routine produces a new estimate of the UCC amplitudes, $\vec{t}^{\ (n+1)}$. The algorithm convergences when the changes in both, total energy and $\vec{t}$, become smaller than suitable thresholds. In the following sections, we describe in detail the steps involved in the VQE implementation of the UCC ansatz. A graphical summary of the procedure is shown in \fig{UCCdiagram}.

\subsection{Implementation of UCC on a quantum computer}





To prepare the UCC ansatz on a quantum computer we need to map the UCC operator (\eq{ucc}) onto operations that can be performed on the quantum computer. We start by rewriting the cluster operator as 
\begin{align}\label{eq:exactU}
U(\vec{t})=e^{\sum_j t_j (\tau_j-\tau_j^{\dagger})}
\end{align}
where $\tau_j$ represent an excitation operator and $t_j$ the corresponding CC amplitude. Since excitation operators do not necessarily commute, the UCC unitary can be approximated using trotterization:
\begin{align}\label{eq:trotterU}
U\left(\vec{t}\right) \approx U_{Trot}\left( \vec{t} \right) = \left( \prod_j e^{\frac{t_j}{\rho}(\tau_j-\tau_j^{\dagger})} \right) ^{\rho}
\end{align}
where $\rho$ is the trotter number. The error associated with the trotter approach depends among other factors, on the norm of the terms being simulated, $\left\| t_j(\tau_j-\tau_j^{\dagger}) \right\|$, which we expect to be small given a reference state with a good overlap with the exact wavefunction. Furthermore, unlike quantum algorithms based on phase estimation, the variational optimization of the parameters in VQE can potentially compensate for the errors associated to the trotterization scheme \cite{Wecker.PRA.92.042303.2015}. In this work we will employ the approximations with $\rho=1$ and $\rho=2$ as our state preparation unitaries. For $\rho=1$:
\begin{align}\label{eq:rho1ucc}
U_{1}\left(\vec{t}\right) \ket{\Phi_{0}} = \prod_j e^{t_j(\tau_j-\tau_j^{\dagger})} \ket{\Phi_{0}}
\end{align}
In the following section we will present numerical evidence that shows that these types of ansatz are as effective as the one in Eq. \ref{eq:exactU}. To implement Eq. \ref{eq:rho1ucc} on a quantum computer, we need to map every unitary in the previous product to operations in the quantum computer. For this purpose we can use either the Jordan-Wigner (JW) or the Bravyi-Kitaev (BK) mappings \cite{jordan.ZP.47.631.1928,Seeley.JCP.137.224109.2012,Tranter.115.1431.IJQC.2015}, obtaining:
\begin{align}\label{eq:transformedExcitation}
(\tau_j-\tau_j^{\dagger})=i\sum^{2^{2l_j-1}}_k P^j_k
\end{align}
where $P^i_j$ represents a product of Pauli matrices with real coefficients and $i$ is the imaginary unit. The index $k$ runs over $2^{2l_j-1}$ products, where $l_j$ is the excitation rank of the j-th excitation operator $\tau_i$ (See \app{appendixA}). We will refer to each $P^j_k$ in Eq.~\ref{eq:transformedExcitation} as a \textit{subterm}. For instance, a double excitation operator minus its complex conjugate will comprise eight subterms. Using the previous notation we can write:
\begin{align}
U_{1}\left(\vec{t}\right)=\prod_j \exp \left(i t_j \sum^{2^{2l_k-1}}_k P^k_j \right)
\end{align}
Furthermore, we can show that the subterms derived from the same $(\tau_j-\tau_j^{\dagger})$ operator commute (See \app{appendixA}), which allow us to simplify the expression of the complex cluster unitary as follows:
\begin{align}\label{eq:finalUCC}
&U_{1}\left(\vec{t}\right)=\prod_j \prod^{2^{2l_k-1}}_k \exp(i t_j P^j_k )
\end{align}
The terms in Eq.~\ref{eq:finalUCC} can be implemented in a quantum computer using the digital model of quantum computation. In this paper we will focus on the universal sets of gates typically employed for superconducting circuit (SQC) and trapped ion (TI) quantum computers \cite{Benhelm.NP.4.463.2008,Barends.N.508.500.2014}: single qubit rotations and CNOT or M{\o}lmer-S{\o}rensen (MS) gates, respectively. 
Thanks to their capabilities in number of qubits and coherent control, the SQC and TI architectures have allowed the first scalable demonstrations of digital quantum simulation \cite{Lanyon.S.334.57.2011,Blatt.NP.8.277.2012,Barends.NC.6.7654.2015}.

Using the first set of gates, the exponentiation of a $n$-fold tensor product of Pauli-Z matrices can be done with $O(N)$ CNOT gates and a single single qubit (SQ) rotation. If there are Pauli-X or Y matrices in the tensor product we must apply the single-qubit Hadamard or $R_x(\frac{\pi}{2})$ gate to rotate to the X or Y basis, respectively, before we compute the parity of the set of qubits with CNOTs, and also apply the inverse gates as part of the uncomputing stage \cite{Whitfield.MP.109.735.2011,Seeley.JCP.137.224109.2012,Tranter.115.1431.IJQC.2015}.

We point out that employing the BK transformation, the number of operations required for implementing a single $\tau_j-\tau_j^{\dagger}$ term scales as $O(\log(N))$ \cite{Seeley.JCP.137.224109.2012}, which represent a most advantageous mapping when compared to the JW transformation that scales as $O(N)$. However, for architectures with limited connectivity (e.g. SQC), we will need extra SWAP operations to implement the exponentiation
, which may eliminate the advantage of the BK transformation. In addition, there is recent evidence that the JW implementation is more robust to errors due to noise in the quantum computer, compared to BK \cite{sawaya2016error}.

The key for retaining a polynomial number of operations to perform VQE with a UCC ansatz is to truncate the CC expansion. A popular truncation in quantum chemistry is to consider only single and double excitations (UCCSD):
\begin{align}
T&\approx T_1+T_2
\end{align}
This approximation suffices to accurately describe many molecular systems and is exact for systems with two electrons. Employing UCCSD, the number of parameters grows as ${N-\eta \choose 2} {\eta \choose 2}+{N-\eta \choose 1} {\eta \choose 1}<O(N^2\eta^2)$ where $N$ is the number of spin orbitals (mapped to qubits) and $\eta$ the number of electrons in the system. Combining the scaling of the number of parameters with upper bounds for the number of gates required to implement a single parameter we can estimate upper bounds for the total number of operations involved in preparing the UCCSD ansatz for single iteration of the VQE algorithm. In the case of the BK transformation, the number of gates scales as $O(N^2\eta^2)$, up to logarithmic factors, compared to $O(N^3\eta^2)$ using the JW transformation. If non-local gates are available (e.g. in TI), the circuit depth for the JW implementation can be reduced by a factor of $O(N)$ using the ordering and parallelization techniques described in \cite{Hastings.IQA.2685188.2015}.

An equivalent alternative to CNOT gates, specifically developed for ion trap architectures, is the M{\o}lmer-S{\o}rensen (MS) gate  \cite{Sorensen.PRL.82.1971.1999,Sorensen.PRA.62.22311.2000}. Its unitary evolution can be represented by the sum over all joint rotations on qubit $j$ and $k$ of the register for an angle $\theta$ around an  axis $\phi$, which can be freely chosen:
\begin{align}
U_\mathrm{MS}(\theta,\phi)=\mathrm{exp}\left(-i\frac{\theta}{2}\sum_{j<k} \sigma_{j}^{\phi} \sigma_{k}^{\phi} \right),
\end{align}
where  $\sigma_{j}^{\phi}=\cos(\phi)\sigma_{j}^{x}+\sin(\phi)\sigma_{j}^{y}$. For $\theta = \pi/2$ and $\phi=0$ the action of $U_\mathrm{MS}$ creates a fully entangled state under $\sigma^x\sigma^x$ operation. This non-local gate can be made to act on  arbitrary subsets of qubits in various ways: (a) by spectroscopic decoupling of unwanted qubits from the interaction \cite{Schindler:2013}, (b) by selectively focussing laser beams on the desired qubits \cite{Debnath:2016} or (c) the use of refocussing techniques \cite{Mueller.NJP.13.85007.2011}.

Depending on the way in which the entangling operations on subregisters are implemented, this leads to a scaling of two entangling operations per parameter, largely reducing their number with respect to the implementation using CNOTs. This is a significant advantage as they remain the limiting factor in the current-day leading architectures, while single qubit operations can already be achieved with very high fidelities far beyond fault-tolerance thresholds. In addition, MS gates are particularly attractive when used with the Bravyi-Kitaev transformation, because the gate only needs to act on $O(\log N)$ qubits rather than $O(N)$ for the Jordan-Wigner transformation.

\subsection{Choice and preparation of the reference state}

In the limit of the complete cluster expansion, the UCC ansatz provides the exact solution for the many body problem. In practice, having a reference state with a high overlap with the exact wavefunction facilitates convergence \cite{Lyakh.CR.112.182.2011}. Generally, the Hartree-Fock solution of the many-body problem provides such reference. The Hartree-Fock state can be written as:
\begin{align}
|\Phi_{0}\rangle = a^{\dagger}_{\eta} a^{\dagger}_{\eta-1} \dots a^{\dagger}_{1} |\rangle
\end{align}
where $|\rangle$ is the fermionic vacuum state. Using the molecular orbital basis, the Hartree-Fock state corresponds to a single product state in the computational basis after the BK or JW mappings are applied. For instance, in the JW mapping the HF state corresponds to the state $|0\rangle^{\otimes N-\eta}\otimes |1\rangle^{\otimes \eta}$, where the the single-particle basis is organized according to the one-particle energy from lowest to highest, the so-called canonical order. In this case the Hartree-Fock state can be constructed by initializing the qubit register with the first $\eta$ qubits in $|1\rangle$ and $N-\eta$ in $|0\rangle$. 

In cases where the molecular wavefunction exhibits strong correlations, the Hartree-Fock state provides a poor starting guess. This problem can be helped by using a multireference approach. One possibility is to employ an entangled reference states obtained from a classical Multiconfigurational Self-Consistent Field (MCSCF) calculation \cite{Szalay.CR.112.108.2011} or a DMRG calculation with a small active space. As long as this state comprises of only a polynomial number of computational states, it can be prepared efficiently on a quantum computer \cite{Ortiz.PRA.64.22319.2001,somma.PRA.65.042323.2002,wang.PRA.79.042335.2009}. Using these reference states, \eq{ucc} can be applied without modification after redefining the space of virtual orbitals according to the occupation of each orbital, which can be determined by measuring the corresponding occupation-number operator. The UCC approach can be also extended to multireference cases by adopting an agnostic unitary coupled cluster ansatz, where the definition of the excitation operators is not linked to a specific reference state, as described in \cite{Mcclean.NJP.18.023023.2016}.

\subsection{Energy measurement}

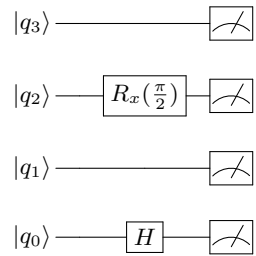
\begin{figure}
\begin{tabular}{c}
\Qcircuit @C=1em @R=1.5em {
|q_3\rangle && \qw  & \qw & \meter \\
|q_2\rangle && \qw  & \gate{R_x(\frac{\pi}{2})} & \meter\\ 
|q_1\rangle && \qw  & \qw & \meter \\
|q_0\rangle && \qw  & \gate{H} & \meter\\
}
\end{tabular}
\caption{Circuit illustrating the measurement of the term $\sigma^{z}_{3}\sigma^{y}_{2}\sigma^{z}_{1}\sigma^{x}_{0}$ in the Z basis. We must apply $H$ or $R_x(-\frac{\pi}{2})$ gates (or equivalent) to change basis when measuring Pauli-Y and Pauli-X operations.}\label{measurePauliStrings}
\end{figure}

Once the state preparation has been performed, the next step in the VQE algorithm is the calculation of the objective function that corresponds to the energy measurement  \mbox{$E= \langle \Phi_{0}| e^{-(T-T^{\dagger})} H e^{T-T^{\dagger}}  |\Phi_{0}\rangle$}. To avoid performing phase estimation, which has a prohibitively large circuit depth for current and near-future quantum devices, we employ the Hamiltonian averaging procedure, introduced in  \cite{McClean.JPCL.5.4368.2014,Mcclean.NJP.18.023023.2016}. In this case the energy is calculated by measuring the expectation value of every term in the Hamiltonian and adding them to obtain the total energy:
\begin{align}\label{eq:decomposedHam}
E=\sum^{M}_{i} h_i \langle O_i\rangle 
\end{align}
where every Hamiltonian term, $O_i$, comprises of a tensor product of Pauli matrices obtained from the JW or the BK transformations, multiplied by the corresponding Hamiltonian coefficient, $h_i$. The expectation value of a string of Pauli matrices, can be measured as illustrated in Figure \ref{measurePauliStrings} using projective measurements. 

We can estimate the number of measurements required to converge the total energy to a precision $\epsilon$ following a frequentist approach, as shown in \cite{Wecker.PRA.92.042303.2015}. Assuming each term in the Hamiltonian is measured $m_i$ times, the precision achieved in each term, $\epsilon_i$, is given by:
\begin{equation}\label{eq:nmeast}
\epsilon_i^2=\frac{|h_i|^2 \text{Var}[\langle O_i \rangle]}{m_i}
\end{equation}
where $\text{Var}[\langle O_i \rangle]$ represents the variance of the expectation value of the operator $O_i$, which is upper-bounded by 1 in the case of Pauli terms. To achieve precision $\epsilon$ in the total energy we can choose $\epsilon_i^2 = \frac{|h_i|}{\sum^{M}_j|h_j|}\epsilon^2$. Taking into account the bound in the variances, we can estimate the total number of measurements, $m$, as:
\begin{equation}\label{eq:nmeas}
m = \frac{\sum^{M}_j |h_j| \sum^{M}_i |h_i|\text{Var}[\langle O_i \rangle]}{\epsilon^2} \le \frac{(\sum^{M}_j |h_j|)^2}{\epsilon^2}
\end{equation}

\subsection{Parameter optimization}

The final step of the VQE algorithm involves the minimization of the total energy with respect to the wavefunction parameters, that in the case of UCC correspond to the cluster amplitudes, $\vec{t}$. This is a non-linear optimization problem for which a variety of optimization algorithms has been proposed \cite{yang2017optimizing}. However, we note that in early demonstration of the VQE algorithm the objective function might exhibit a highly non-smooth character due to experimental noisy conditions. In this scenario, we might expect that direct search algorithms, which are more robust to noise, have an advantage over optimization methods that rely on gradients \cite{kolda.SIAMR.45.385.2003}.

The optimization performance will also depend on the quality of the starting parameters. Fortunately, it is possible to generate starting guesses for the cluster amplitudes based on classical quantum chemistry approaches. For instance, classical CCSD employ the CC amplitudes obtained from second order M{\o}ller-Plesset perturbation theory (MP2) as starting guesses to solve for the CC equations. The MP2 guess amplitudes are given by the equations:
\begin{align}
t^a_i= 0; \quad t^{ab}_{ij} = \frac{h_{ijba} - h_{ijab} }{\epsilon_i + \epsilon_j - \epsilon_a - \epsilon_b }    
\end{align}
where $\epsilon_p$ stands for the Hartree-Fock energy of the orbital $p$ and $h_{pqrs}$ represent the two electron integrals (\eq{double_int}). This  information is obtained directly from the Hartree-Fock calculation. As the solutions of truncated CC or truncated CI are also efficient, it is possible to use cluster amplitudes obtained from methods such as CCSD. One can easily compute both cluster amplitudes and MP2 amplitudes using modules provided in OpenFermion \cite{openfermion}.

Classical approximations to the cluster amplitudes also serve as a criteria to reduce the number of parameters in the optimization. Before starting the VQE optimization, we can remove from the UCC unitary those excitation operators that have a small amplitude according to the classical estimate, as they are likely to also have a small contribution to the final wavefunction. Once the first optimization has been completed, we might include more excitation operators and repeat the optimization until a desired convergence threshold is achieved. The same strategy could be employed during the optimization process, discarding those operators for which the cluster amplitudes remain small after certain number of VQE iterations.

\begin{figure*}
\begin{tabular}{c}
\Qcircuit @C=1em @R=1.5em {
&&& |0\rangle && \qw & \qw & \gate{H} & \ctrl{1} & \qw & \qw & \qw & \qw & \ctrl{1} & \gate{H} & \gate{R_x(\frac{\pi}{2})}&\meter \\
|\Phi_0\rangle && {/} \qw & \gate{e^{i t_1 P^1_{1}}} &  \qw & \cdots && \gate{e^{i t_j P^j_{k}}} & \gate{P^j_k} & \gate{e^{i t_j P^j_{k+1}}} &\qw  & \cdots && \gate{O_i} & \qw & \qw \\ 
}
\end{tabular}
\caption{Circuit for measuring the imaginary part of  $\langle \Phi_0| V^{j\dagger}_{k}(\vec{t}) O_i  U(\vec{t})|\Phi_0\rangle$ required in the calculation of the partial derivative $\frac{\partial E(\vec{t})}{\partial t_j}$. The $R_x(\frac{\pi}{2})$ gate rotates to the $Y$-basis.}\label{fig:fig2}
\end{figure*}
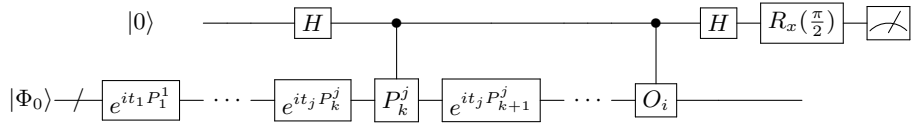

\subsection{Gradient evaluation for UCC}\label{sec:UCCderivatives}

Direct search algorithms can be more robust to noise than gradient-based approaches, but this generally comes at the cost of demanding a larger number of function evaluations to achieve convergence \cite{kolda.SIAMR.45.385.2003}. As the accuracy of quantum computers increases, the possibility of computing energy gradients in the quantum computer becomes more feasible. One possibility is to compute the gradient numerically, using for instance a finite difference formula. In this case, the accuracy of the gradient depends on the step size chosen, which would be limited by the precision of the experimental control over the parameters and by shot-noise limited measurements. 

Alternatively, one might evaluate the gradient directly on the quantum computer given that an analytical implementation is available. Here we propose a method to compute the analytical gradient of the energy when a product of parametrized unitaries is employed in the state preparation. 

Consider a unitary ansatz analogous to the one defined in \eq{finalUCC}:
\begin{align}\label{eq:finalUCC2}
&U \left(\vec{t}\right) = \prod^{N_P}_j \prod^{N^j_S}_k \exp(i c^j_k t_j P^j_k )
\end{align}
where $N_P$ stands for the number of parameters and $N^j_S$ stands for the number of subterms that depend on the $j$-th parameter. $P^j_k$ is a string of Pauli matrices. $c^j_k$ is a constant, that in the case of the UCC ansatz corresponds to the constant factors obtained arising from the mapping of fermionic operators to qubit operators. Consider the state $\Psi\left(\vec{t}\right)$, prepared as $\Psi\left(\vec{t}\right) = U\left(\vec{t}\right) |\Phi_0\rangle$, where $|\Phi_0\rangle$ is a reference wavefunction that do not depend on $\vec{t}$. Also consider a Hamiltonian, $H$, which is independent of the parameters $\vec{t}$. In this case, the derivative of the expectation value of the energy, $E(\vec{t})=\langle \Psi\left(\vec{t}\right) | H |  \Psi\left(\vec{t}\right)\rangle$, with respect to the parameter $t_j$ will be given by
\begin{align}\label{eq:exactDerivative}
&\frac{\partial E(\vec{t})}{\partial t_j} = \langle \Phi_0|U^{\dagger}(\vec{t}) H  \frac{\partial U(\vec{t})}{\partial t_j} |\Phi_0\rangle + \langle \Phi_0| \frac{\partial U(\vec{t})^{\dagger}}{\partial t_j} H  U(\vec{t})|\Phi_0\rangle \\
& = i\sum^{N^{j}_S}_{k} \langle \Phi_0|U^{\dagger}(\vec{t}) H  V^{j}_{k}(\vec{t}) |\Phi_0\rangle - \langle \Phi_0| V^{j\dagger}_{k}(\vec{t}) H  U(\vec{t})|\Phi_0\rangle \\
\label{eq:derivative}
& = 2 \sum^{N^{j}_S}_{k} c^j_k \operatorname{Im} (\langle \Phi_0| V^{j\dagger}_{k}(\vec{t}) H  U(\vec{t})|\Phi_0\rangle)
\end{align}
Where the operator $V^{j}_{k}(\vec{t})$ is defined as the unitary of \eq{finalUCC2} but with the operator $P^{j}_{k}$ interleaved between the unitaries $\exp(i t_j P^j_{k-1} )$ and $\exp(i t_j P^j_k )$. Explicitly:
\begin{align}
V^{j}_{k}(\vec{t}) = & \exp(i t_j P^1_1) \cdots \exp(i t_j P^j_{k-1}) P^j_{k} \exp(i t_j P^j_{k}) \notag \\ & \exp(i t_j P^j_{k+1}) \cdots \exp(i t_{N_P} P^{N_P}_{N^{N_P}_S})
\end{align}
Combining \eq{derivative} with the decomposition of the Hamiltonian in \eq{decomposedHam}, we obtain a working expression for computing $\frac{\partial E(\vec{t})}{\partial t_j}$:
\begin{align}\label{eq:finalderivative}
\frac{\partial E(\vec{t})}{\partial t_j} & = 2 \sum^{M}_{i} h_i \left( \sum^{N^{j}_S}_{k} c^j_k \operatorname{Im} (\langle \Phi_0| V^{j\dagger}_{k}(\vec{t}) O_i  U(\vec{t})|\Phi_0\rangle) \right)
\end{align}
We can evaluate the imaginary part of $\langle \Phi_0| V^{j\dagger}_{k} H_i  U(\vec{t})|\Phi_0\rangle$ with the circuit of \fig{fig2}. Here, we use a state register initialized with the reference state tensor an ancilla qubit initialized in a superposition. First, we apply the unitaries of \eq{finalUCC2} to the state register up to $\exp(i t_j P^j_k )$, after which we apply the operator $P^j_k$ controlled by the ancilla qubit. Subsequently, we apply the remaining unitaries to the state register, followed by the local operator $H_i$ controlled by the ancilla qubit. Finally, we apply a Hadamard gate in the ancilla qubit to obtain the state
\begin{align}
\frac{\ket{0}\otimes(U\ket{\Phi_0}+ O_i V^j_k(\vec{t})\ket{\Phi_0}) + \ket{1}\otimes(U\ket{\Phi_0} - O_i V^j_k(\vec{t})\ket{\Phi_0})}{2}
\end{align}
The imaginary part of $\langle \Phi_0| V^{j\dagger}_{k}(\vec{t}) O_i  U(\vec{t})|\Phi_0\rangle$ can be recovered by measuring the ancilla qubit in the $Y$-basis.
\jrf{The variance of the $j$-th component of the gradient as computed with the circuit of \fig{fig2} will be given by:
\begin{equation}\label{eq:gradient_variance}
\text{Var} \left[ \frac{\partial E(\vec{t})}{\partial t_j} \right] = 4 \sum^{M}_{i} |h_i|^2 \sum^{N^j_S}_{k} |c^j_k|^2 \text{Var}\left[\langle \sigma^y \rangle_{O_i, P^j_k} \right] 
\end{equation}
where
\begin{equation}\label{eq:var_sigmay}
\langle \sigma^y \rangle_{O_i, P^j_k} = \left \langle 0 \otimes \Phi_0 \left| C_{O_i, P^j_k}^{\dagger} (\sigma^y \otimes I) C_{O_i, P^j_k} \right| 0 \otimes \Phi_0 \right \rangle
\end{equation}
and $C_{O_i, P^j_k}$ represents the circuit for gradient estimation for the subterm $P^j_k$ and the observable $O_i$. To estimate the number of measurements required to achieve precision $\tilde{\epsilon}_j$ in the $j$-th component of the gradient, we will first consider the number of measurements required to estimate the contribution of the circuit $C_{O_i, P^j_k}$ to precision $\tilde{\epsilon}^i_{j, k}$:
\begin{equation}\label{eq:nmeas_subterm_grad}
\tilde{m}^{i}_{j,k} = \frac{|c^j_k|^2 \text{Var} \left[\langle \sigma^y \rangle_{O_i, P^j_k} \right]}{(\tilde{\epsilon}^i_{j, k})^2} 
\end{equation}
For the UCC ansatz, the constants $c^{i}_{j, k}$ have the same norm, $|c^{i}_{j, k}|=|c_{j}|$ and fulfill $\sum^{N^{j}_{S}}_{k} |c^{i}_{j, k}| = 1$. Therefore we can choose $(\tilde{\epsilon}^i_{j, k})^2 = |c_{j}| (\tilde{\epsilon}^i_j)^2$, where $\tilde{\epsilon}^i_j$ is the precision for estimating the contribution of the operator $O_i$ to the gradient variance. In addition, we can apply the same sampling strategy chosen for estimating the energy (\eq{nmeas}), and choose $(\tilde{\epsilon}^i_j)^2 = \frac{|h_i| \tilde{\epsilon}_j^2}{\sum^{M}_l |h_l|}$. Introducing these considerations into \eq{nmeas_subterm_grad}, we obtain:
\begin{equation}\label{eq:nmeas_grad}
\tilde{m}_{j} = \left( 4 \sum^{M}_{l} |h_l| \right) \frac{\sum^{M}_{i} \sum^{N^{j}_{S}}_{k}  |h_i| |c^j_k| \text{Var} \left[\langle \sigma^y \rangle_{O_i, P^j_k} \right]}{\tilde{\epsilon}^2_j} 
\end{equation}
We can get an upper bound to \eq{nmeas_grad} by considering the upper bound of the variance and including the properties of the coefficients $c^{i}_{j, k}$:
\begin{equation}\label{eq:upper_nmeas_grad}
\tilde{m}_{j} \le 4 \frac{\left(\sum^{M}_{i} |h_i| \right)^2}{\tilde{\epsilon}^2_j} 
\end{equation}
For comparison, consider the simplest central finite difference formula that requires two energy evaluations to compute each gradient component:
\begin{equation}\label{eq:centraldiff}
\frac{\partial E(\vec{t})}{\partial t_j} \approx \frac{E(t_1,..,t_j+\delta,..,t_{N_P})-E(t_1,..,t_j-\delta,..,t_{N_P})}{2\delta}
\end{equation}
where $\delta$ is the step size. As in the case of the analytical gradient, we choose to estimate the $j$-th gradient component to precision $\tilde{\epsilon}_j$. The precision in the numerical gradient will depend on the precision of the numerator and denominator of \eq{centraldiff}. Assuming no error in the denominator and a non-zero numerator, the precision for estimating the energies in the numerator, $\epsilon_j$, can be chosen to guarantee that the relative precisions of the gradient component and the numerator are the same. This condition requires $\epsilon_j = \frac{2\delta\tilde{\epsilon}_j}{\sqrt{2}}$. Combining this requirement with \eq{nmeas}, we can bound the number of measurements for estimating the $j$-th component of the gradient as:
\begin{equation}\label{eq:nmeasgn}
\tilde{m}_{j} \le 4 \left( \frac{(\sum^{M}_{i} |h_i|)^2}{(2\delta)^2 \tilde{\epsilon}_j^2}\right),
\end{equation}
where the estimate considers two energy evaluations per gradient component. To achieve precision $\tilde{\epsilon}$ in the norm of the gradient, we  could choose $\tilde{\epsilon}^2_j = \frac{\tilde{\epsilon}^2}{N_P}$, allowing us to bound the sampling cost of gradient approximations as:
\begin{equation}\label{eq:approx_nmeas_grad}
\tilde{m} \le C N_P\left( \frac{(\sum^{M}_{i} |h_i|)^2}{\tilde{\epsilon}^2}\right),
\end{equation}
where $C=\frac{4}{(2\delta)^2}$ for the simplest central difference formula and $C=4$ for the analytical gradient estimated using \fig{fig2}. The same bounds can be derived for the UCC approximations with more than one Trotter step, $\rho>1$. In this case, the factor $\frac{1}{\rho}$ appears multiplying the constants $c^{i}_{j, k}$, but the number of circuits contributing to $N^{j}_{S}$ also increases by factor of $\rho$, canceling out the $\frac{1}{\rho}$ factor in the estimation of the bound.

The previous analysis indicates that the sampling cost of the numerical gradient increases quadratically with decreasing the step size. From the analysis of \eq{approx_nmeas_grad}, we expect that for $\delta < 0.5$ the numerical gradient will have a larger sampling cost than the analytical gradient approach. In addition, the accuracy of the numerical gradient depends on the step size used in the central difference formula and sets a lower bound to the precision that can be obtained from the numerical gradient. 

From \eq{approx_nmeas_grad}, we also conclude that the gradient estimation is more expensive than estimating the energy by a factor proportional to the number of parameters. However, the relative cost of gradient-based and gradient-free optimization is ultimately determined by the number of iterations required for convergence. Usually, gradient based methods employ a number of gradient evaluations much smaller than the number of energy evaluations employed by derivative-free methods.

Finally, we point out that the sampling cost can be reduced by adapting the precision required in each optimization step according to the norm of the gradient, instead of  employing a fixed gradient precision throughout the optimization. With this strategy, the first steps would require less measurements compared to the final steps, where the gradient norm is smaller.}

\subsection{VQE-UCC with an active space approximation}\label{theory:approximateMethods}

Several approximations that have been designed to reduce the computational cost of classical quantum chemistry algorithms can be extrapolated to the quantum implementation. A particular strategy that could be exploited to reduce the number of quantum resources for a VQE-UCC calculation is the \textit{complete active space} (CAS) approach \cite{Roos.CP.48.157.1980}. The CAS approximation consists in dividing the orbital space into a set of \textit{inactive} ($I$) and \textit{active} ($A$) orbitals such as the occupation of the orbitals in the inactive space remains unchanged. This idea exploits the fact that for most of the quantum chemistry Hamiltonians, including those cases with a strong multireference character, the wavefunction is qualitatively dominated by a relatively small number of Slater determinants that can be effectively captured by expanding the wavefunction in a subspace defined by the active orbitals. 

In most quantum chemistry applications, the CAS approximation is employed to treat static correlation effects, meaning that a relatively small active space is selected to obtain a qualitatively correct wavefunction that serves as reference state for further perturbation theory or Coupled Cluster refinements \cite{Lyakh.CR.112.182.2011,Szalay.CR.112.108.2011}. Nonetheless, one might also consider the choice of an active space that is sufficiently large such as both static and dynamical correlation effects can be described up to certain accuracy.

In the case of the CAS approximation applied to single reference UCC, one selects an active space comprised of $\eta_{A}$ electrons distributed among $N_A$ spatial orbitals. This choice of active space is denoted as $CAS(\eta_{A}, N_A)$. The active orbitals usually correspond to a selection of the highest occupied orbitals and the lowest virtual orbitals. The cluster operators are then redefined such as excitations are only allowed among active orbitals,
\begin{align}\label{cas:T}
T^A&=\sum^{\eta_A}_{i} T_i
\end{align}
Considering the separation between active and space orbitals, we can rewrite the reference state as $|\Phi_{0}\rangle= |\Phi^{A}_{0}\rangle \otimes |\Phi^{I}_{0}\rangle$, where $|\Phi^{I}_{0}\rangle$ and $|\Phi^{A}_{0}\rangle$ are defined over the inactive space and active space, respectively. Consequently we can write the total energy as:
\begin{align}
\label{effectiveUCCenergy}
E= \langle \Phi^{A}_{0}| e^{-(T^{A}-T^{A \dagger})} \tilde{H}^{A} e^{T^{A}-T^{A\dagger}}  |\Phi^{A}_{0}\rangle
\end{align}
where the effective Hamiltonian $\tilde{H}^{A}$ is obtained by evaluating the following expression:
\begin{align}\label{effectiveH}
\tilde{H}^{A}= \langle \Phi^{I}_{0}|H|\Phi^{I}_{0}\rangle
\end{align}
Since the reference state corresponds to a product state, the calculation of the effective Hamiltonian can be performed efficiently on a classical computer. In this case, we can obtain an approximate solution to the VQE problem by performing a VQE-UCC calculation for the effective active space Hamiltonians, $H_{jj'}^{A}$. The CAS-UCC approach reduces the number of qubits required for a calculation by a factor of $N_A/N$. Similarly, the number of parameters for the preparation of the UCCSD wavefunction is reduced by a factor of $(\eta_A N_A)^2/(N\eta)^2$ with respect to full-UCCSD, as the scaling becomes $O({\eta_{A}}^2 N_{A}^2)$.

A number of strategies for selecting active spaces to describe static correlation have been proposed in the context of quantum chemistry. Generally, these strategies employ the occupation of approximate natural orbitals, which are the orbitals that diagonalize the one particle density matrix, as a criteria to choose the active space. Orbitals with integer occupation are generally discarded, and only those with fractional occupation within certain thresholds are considered as part of the active space. The approximate one particle density matrix is obtained from methods that include some amount of correlation and that are relatively  inexpensive, such as MP2 \cite{jensen.JCP.88.3834.1988}. Another commonly used approach employs the unrestricted natural orbitals (UNO) obtained from unrestricted Hartree-Fock calculations \cite{abrams.CPL.395.227.2004,keller.JCP.142.244104.2015}. More recently, a scheme based on entanglement measurement among orbitals has been also proposed \cite{stein.JCTC.12.1760.2016}.

We can take advantage of one of these approaches to define an initial active space in a suitable basis for the UCC calculation. The solution obtained with the initial active space can be employed as an initial guess for another CAS-UCC calculation with a larger active space. This process can be repeated until the simulation is performed on the entire basis, in which case we expect the algorithm to converge faster as in each iteration a better approximation to the exact UCC wavefunction is obtained. One can also stop the optimization after the energy does not improve beyond a pre-defined threshold. In the later case, we also achieve a reduction in the number of qubits required for the calculation.

\section{Numerical assessment}\label{sec:numerics}

\subsection{Classical simulation of VQE-UCC}


\jrf{To illustrate the algorithmic details of the the scalable VQE-UCC algorithm, we simulated the VQE-UCC calculation of small molecules. The molecular integrals were obtained using the PSI4 package \cite{Schmidt.JCC.14.1347.1993} and the molecular Hamiltonian was mapped using the Jordan-Wigner transformation. The UCC unitary was constructed with a truncated cluster operator and the symbolic representation was transformed into unitaries comprising strings of Pauli matrices, following the same procedure employed for the Hamiltonian. To assist these transformations, we employed the OpenFermion (\url{www.openfermion.org}) library \cite{openfermion}.}

The simulation of the circuit proceeds by calculation of the UCC wavefunction from the the matrix representation of the UCC unitary and the reference state. The optimization was performed using three direct search algorithms available in the \emph{scipy.optimize} library, namely the Nelder-Mead \cite{neldermead}, Powell \cite{powell} and COBYLA \cite{cobyla} algorithms. We also employed the L-BFGS-B method \cite{byrd.SIAMJSC.16.1190.1995} with numerical gradients for comparison, using the central finite difference formula (\eq{centraldiff}). The energy and parameter thresholds for convergence were fixed at 10$^{-5}$ a.u 10$^{-4}$ a.u respectively. For the L-BFGS-B algorithm, the convergence threshold for the projected gradient was fixed at 10$^{-4}$. In all cases the maximum number of function evaluations was fixed to 20,000. Finally, we point out that all our numerical experiments assume that function evaluations are performed in double precision arithmetic, unless indicated otherwise.


\subsection{VQE-UCC results for H$_4$ molecular systems}

A practical and informative assessment of quantum chemistry simulation involves the study of chemical transformations, such as bond-breaking, isomerization or configurational changes. These processes are generally described through scanning geometries along certain directions of a PES. Along the PES, the amount of entanglement of the wavefunction varies greatly and this impacts the performance of the ansatz employed to approximate the wave function. 

\begin{figure}
\centering
\includegraphics[width=8cm]{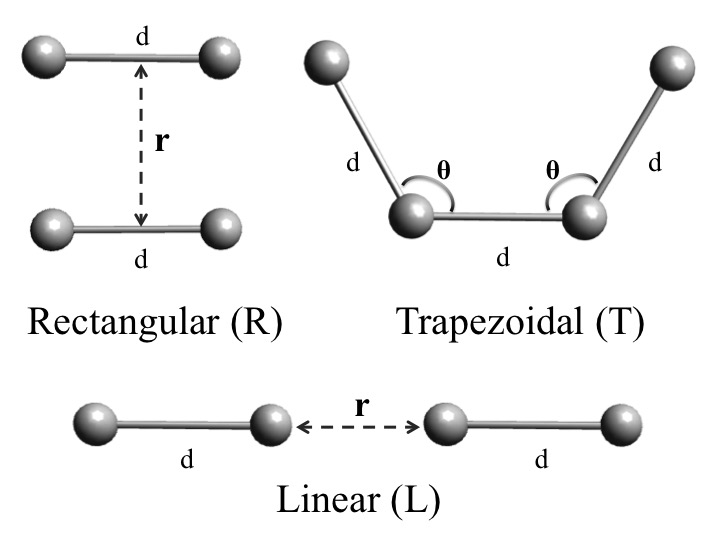}
\caption{Description of geometries for the H$_4$ model systems studied in this work. The potential energy surfaces are defined as a function of the variable $r$ for the rectangular (R) and linear (L) geometries and as a function of $\theta$ for the trapezoidal (T) geometry. The value of the parameter d is kept fixed at 2.0 \AA.}\label{fig:H4models}
\end{figure}

In order to illustrate these aspects, we selected a model in which the amount of entanglement in the wavefunction can be continuously varied and which is simple enough to enable simulations. We have considered the PES of a system comprising four hydrogen atoms investigated along three different paths: rectangular (R), trapezoidal (T), and linear (L), as described in \fig{H4models}. These systems have been widely employed by the quantum chemistry community as a benchmark for multireference methods  \cite{jankowski.IJQC.18.1242.1980,Lyakh.CR.112.182.2011}. We studied 19 different geometries for the trapezoidal path generated by varying the parameter $\theta$ between $90\degree$ and $180\degree$. For the linear and the parallel paths, we studied 24 different geometries generated by varying the parameter $r$ between 0.6{\AA} and 5.0{\AA}.

\subsubsection{Influence of the optimization method in the VQE performance}

We evaluated the effectiveness of the strategies proposed to generate the initial guess for the cluster amplitudes and optimization methods based on three criteria: \begin{enumerate} 
\item the error in the calculated energy with respect to the FCI solution, \mbox{$E_{\mathrm{FCI}}-E_{\mathrm{VQE}}$}, 
\item the accuracy of the wavefunction evaluated as the infidelity ($1-|\langle \Psi_{\mathrm{VQE}} | \Psi_{\mathrm{FCI}}\rangle|^2$) and 
\item the number of function evaluations required for convergence. \end{enumerate} 
We compared the four optimization methods described in the previous section using three different starting guesses: \begin{enumerate} 
\item random, in which random values are chosen uniformly in the interval -0.25 to 0.25, 
\item starting with all the amplitudes set to zero, which corresponds to using the Hartree-Fock solution as initial guess and
\item the MP2 approximation to the cluster amplitudes. \end{enumerate} 
The full optimization is comprised of a total of 52 parameters. To evaluate the performance of the random guess approach we ran the algorithm 10 times and averaged the results.

\begin{figure*}
\centering
\begin{tabular}{cc} 
a) & \includegraphics[width=10cm]{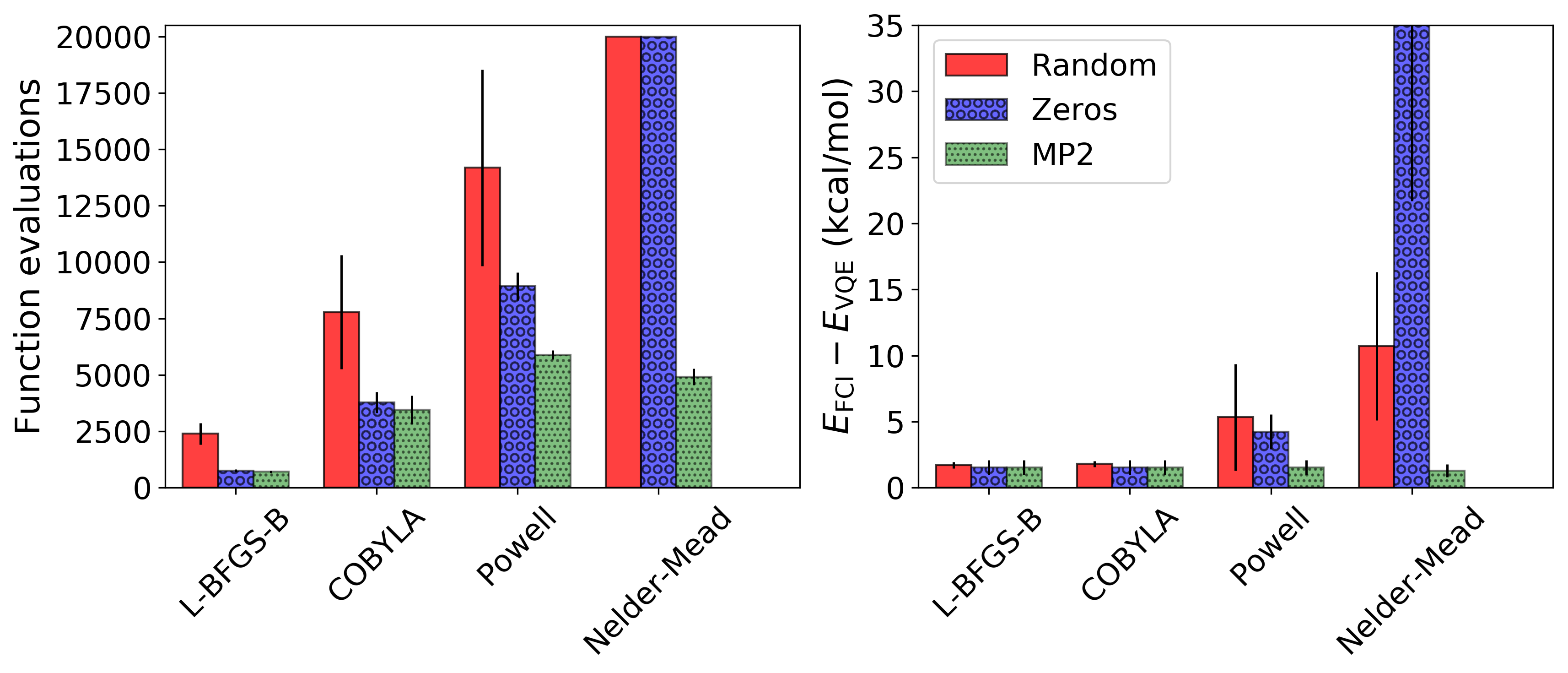} \\
b) & \includegraphics[width=10cm]{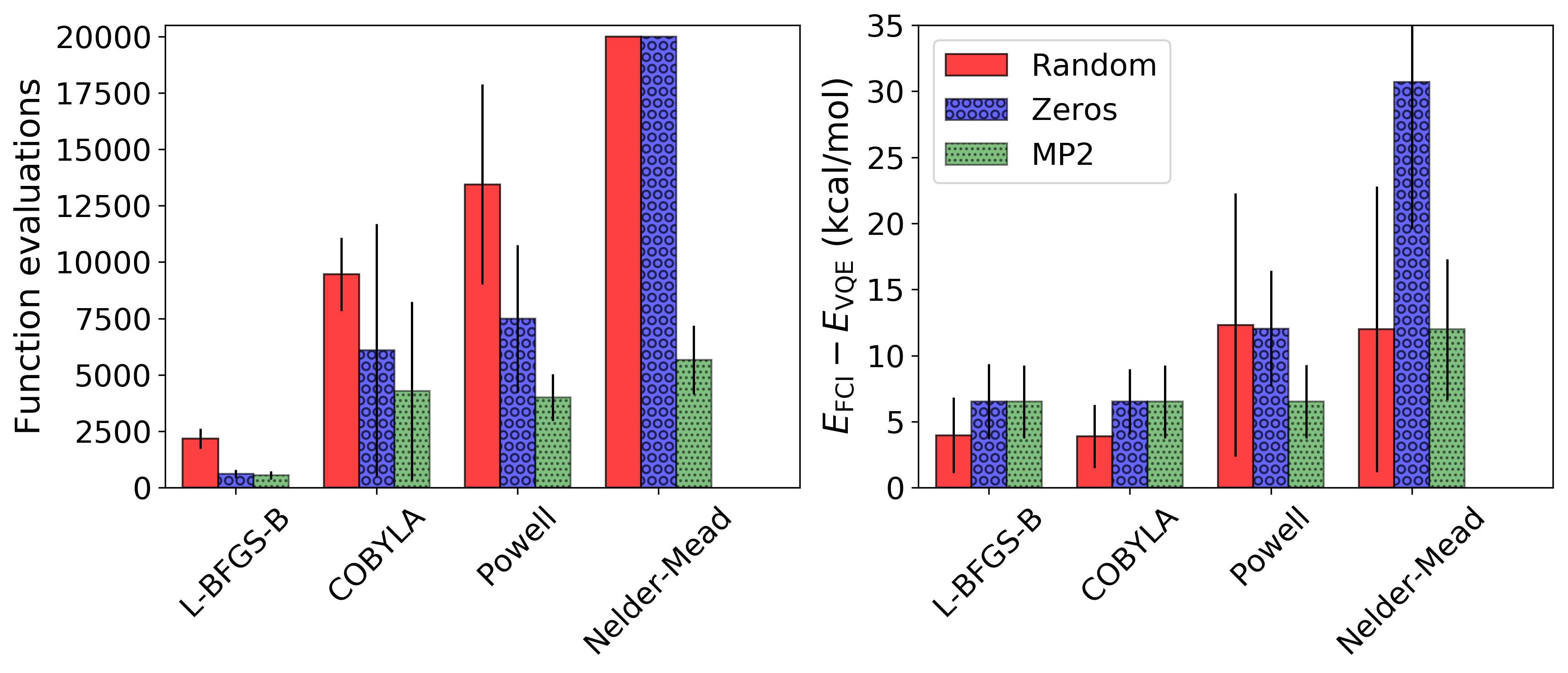} \\
c) & \includegraphics[width=10cm]{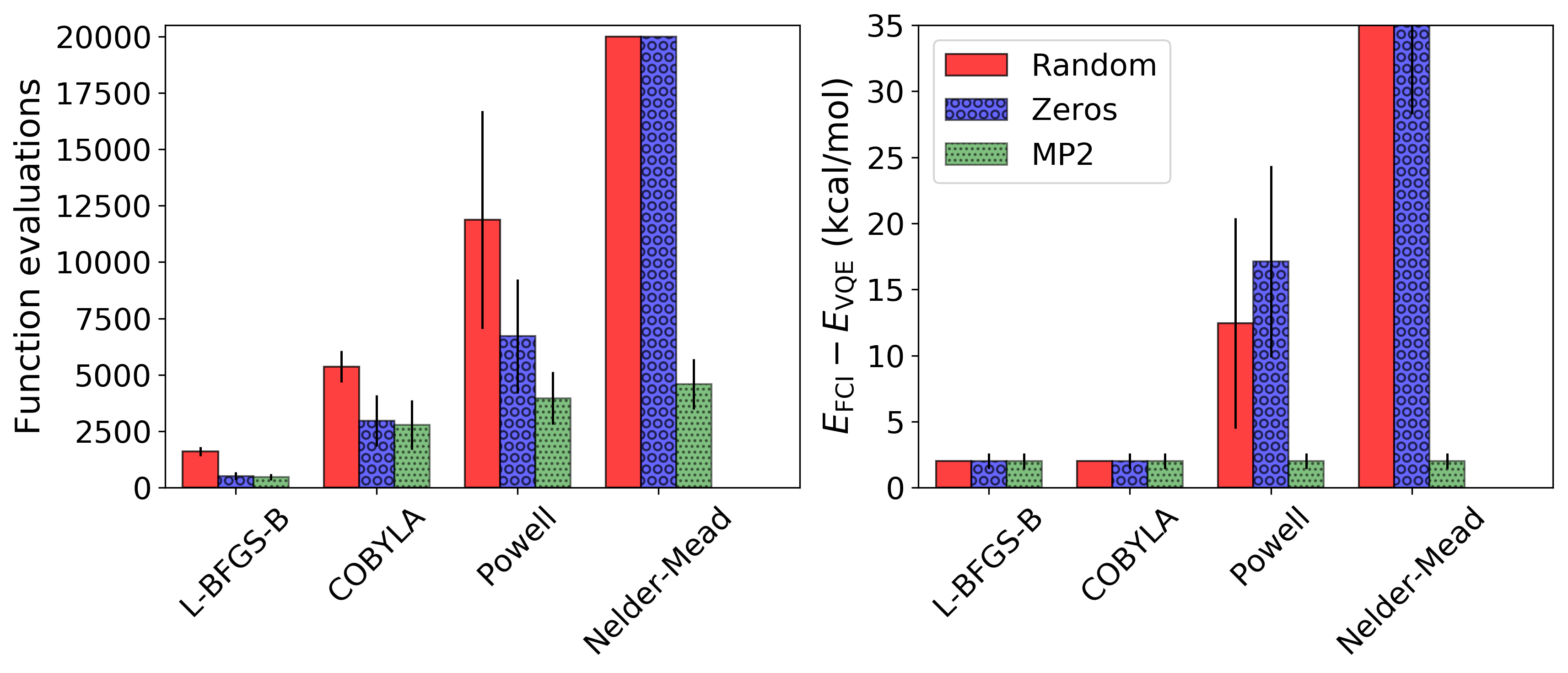} \\
\end{tabular}
\caption{Average performance of the VQE algorithm applied to the H$_4$ system along the a) trapezoidal b) linear and c) parallel paths using four optimization methods (L-BFGS-B, COBYLA, Powell and Nelder-Mead) and three different methods to initialize the parameters: Randomly (Random), set to zero (Zeros) and set to the MP2 amplitudes (MP2). We compare the number of function evaluation required for convergence (left panel) and the error in the final energy with respect to the FCI solution (right panel). Error bars indicate one standard deviation. The range in the energy plots is truncated to 35 kcal/mol to facilitate comparison.}\label{fig:optimizationT}
\end{figure*}

\fig{optimizationT} compares the average number of function evaluations and energy error along the rectangular, trapezoidal, and linear paths of the $\rm{H}_4$ system. We observe that the Nelder-Mead and the Powell methods exhibit a high variability in their performances when the parameters are initialized at zero or randomly, as indicated by the large standard deviations in the wavefunction accuracy. In particular, Nelder-Mead fails to converge in less than 20000 function evaluations and performs poorly, with energy errors beyond 10~kcal/mol and overlaps with the exact wavefunction below 0.8. The Powell method has a better performance in the number of function evaluations but is still outperformed by L-BFGS and COBYLA. On the other hand, the COBYLA and the L-BFGS-B methods converge to almost the same minimum for most of the points of the PES, independent of the method employed to generate the initial guess. This is indicated by the much larger energy accuracies compared to the results of Nelder-Mead and Powell.

The use of the MP2 guesses for the cluster amplitudes significantly reduces the number of function evaluations for all the optimization methods as observed in the left panel of \fig{optimizationT}. MP2 guesses also improve the average accuracy of the energy obtained with the Nelder-Mead and Powell methods, as observed in the right panel of \fig{optimizationT}. We point out that for systems that experience strong correlation the MP2 amplitudes might be a poor starting point, although still better than the random or zeros guesses. In those cases, more reliable methods such as Density Matrix Renormalization Group (DMRG) with a small active space and a small bond dimension could provide better initial guesses at the expense of classical computation time \cite{veis.JPCL.7.4072.2016}. These results illustrate how the incorporation of classical approaches can improve the performance of quantum simulation by providing physically meaningful starting guesses and also highlight the importance of the choice of the optimization method. 

\subsubsection{Effect of trotterization in the optimization}

\tab{trotterization} compares the performance of the trotterized UCC ansatz (\eq{trotterU}) using 1 and 2 trotter steps with the performance of the non-trotterized ansatz (\eq{exactU}). For these calculations we employed the COBYLA and the L-BFGS-B optimization methods with the MP2 guess. To measure the quality of the results we use the average infidelity with respect to the FCI wavefunction as well the non-paralellism error (NPE). The NPE is calculated according to the formula:
\begin{align}\label{NPE}
\mathrm{NPE} = \max(E_\mathrm{\rm UCCSD}-E_\mathrm{FCI})-\min(E_\mathrm{UCCSD}-E_\mathrm{FCI})
\end{align}
which quantifies the maximum error obtained when computing energy differences between points in the PES using the UCCSD approach. As observed in \tab{trotterization}, the quality of the results obtained with the trotterized unitaries is almost identical to the that of the exact implementation of \eq{exactU} when using the L-BFGS-B optimization method. We also notice that the approximation with 2 trotter steps converges faster in average than the unitary with only one trotter step. 

\begin{figure*}
\centering
\begin{tabular}{cc}  \includegraphics[height=5cm]{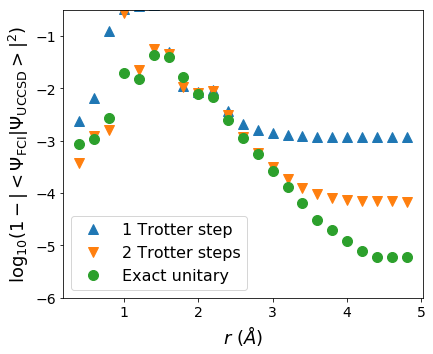} & \includegraphics[height=5cm]{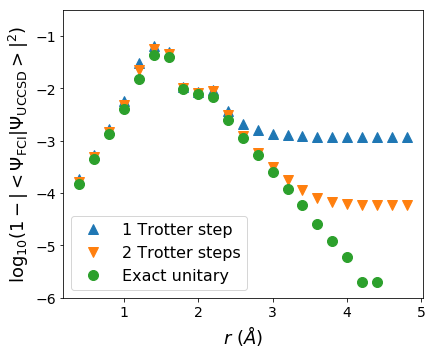} \\
(a) & (b) \\
\end{tabular}
\caption{Comparison between the error in the wavefunctions obtained using a) COBYLA and b) L-BGFG-B optimizations along the linear path of the $\rm{H}_4$ system. The UCCSD wavefunctions were obtained using the exact UCC unitary (\eq{exactU}, red dots) and the trotterized version (\eq{finalUCC}) with one trotter step (blue triangles up) and two trotter steps (green triangles down). The error in the wavefunction is quantified as $1-|\langle \Psi_{\mathrm{UCCSD}} | \Psi_{\mathrm{FCI}}\rangle|$.}\label{fig:resultstrotterLinear}
\end{figure*}

Using COBYLA, the trotterized unitaries produce results similar to those obtained with \eq{exactU} for the trapezoidal and the parallel paths. In contrast, COBYLA exhibits a lower average performance for the linear system, as shown in \tab{trotterization}. A better insight into this result is offered by \fig{resultstrotterLinear}, where we plot the error in the wavefunction along $r$ for the linear path as computed with the COBYLA and the L-BFGS-B methods. The error in the wavefunction is quantified as the difference between 1.0 and the absolute value of the overlap of the UCCSD and the FCI wavefunctions. We observe that the COBYLA algorithm provides wavefunctions with overlaps below $0.95$ and as low as $0.78$ between $0.8-1.6$~{\AA}, which corresponds to a section of the PES with strong multireference character. For these geometries, the COBYLA algorithm reaches the maximum number of functions evaluations when using \eq{trotterU} with 1 trotter step. Increasing the number of trotter steps seems to partially alleviate this problem. In contrast, a gradient based approach such as L-BFGS-B provides better results in the $0.8-1.6$~{\AA} range. Interestingly, as the distance increases beyond $2.6$~{\AA}, the difference between the trotterized and the exact unitary becomes more prominent. We point out, however, that in all these cases the overlap is larger than $0.999$ with a single trotter step.

\begin{table*}[h]
\caption{Comparison of the VQE results obtained with the ansatz from \eq{exactU} and \eq{finalUCC} with one and two trotter steps, for the calculation of the PES of the H$_4$ systems. We compared the average overlap with the FCI wavefunction, non-parallelism error and average number of function evaluations along the trapezoidal, parallel and linear paths of the H$_4$ system, obtained using the L-BFGS-B and COBYLA optimization methods. The molecular Hamiltonian was obtained with a STO-6G basis set.}\label{tab:trotterization}
\begin{tabular}{cccccc}
\hline
\specialcell{Optimization\\method} & System & Approximation & \specialcell{Average Overlap} & \specialcell{NPE in PES$^*$\\(kcal/mol)} & \specialcell{Number of energy\\evaluations}\\
\hline
L-BFGS-B	&	Trapezoidal	&	\eq{finalUCC} $\rho=1$	&	0.994 $\pm$ 0.006	&	1.4	&	861 $\pm$ 73	\\
	&		&	\eq{finalUCC} $\rho=2$	&	0.995 $\pm$ 0.005	&	1.5	&	615$\pm$ 32	\\
	&		&	\eq{exactU}	&	0.995 $\pm$ 0.005	&	1.5	&	703 $\pm$ 51	\\
\hline											
	&	Parallel	&	\eq{finalUCC} $\rho=1$	&	0.996 $\pm$ 0.008	&	2.0	&	590 $\pm$ 144	\\
	&		&	\eq{finalUCC} $\rho=2$	&	0.997 $\pm$ 0.007	&	2.0	&	436 $\pm$ 149	\\
	&		&	\eq{exactU}	&	0.997 $\pm$ 0.006	&	2.0	&	467 $\pm$ 142	\\
\hline											
	&	Linear	&	\eq{finalUCC} $\rho=1$	&	0.998 $\pm$ 0.006	&	7.1	&	710$\pm$ 99	\\
	&		&	\eq{finalUCC} $\rho=2$	&	0.999 $\pm$ 0.005	&	6.9	&	487 $\pm$ 158	\\
	&		&	\eq{exactU}	&	0.999 $\pm$ 0.005	&	6.5	&	534 $\pm$ 182	\\
\hline											
COBYLA	&	Trapezoidal	&	\eq{finalUCC} $\rho=1$	&	0.994 $\pm$ 0.006	&	1.0	&	3703 $\pm$ 1023	\\
	&		&	\eq{finalUCC} $\rho=2$	&	0.995 $\pm$ 0.005	&	1.5	&	2753 $\pm$ 340	\\
	&		&	\eq{exactU}	&	0.995 $\pm$ 0.005	&	1.5	&	3468 $\pm$ 622	\\
\hline											
	&	Parallel	&	\eq{finalUCC} $\rho=1$	&	0.998 $\pm$ 0.006	&	2.0	&	2431 $\pm$ 857	\\
	&		&	\eq{finalUCC} $\rho=2$	&	0.999 $\pm$ 0.005	&	2.0	&	2047 $\pm$ 665	\\
	&		&	\eq{exactU}	&	0.999 $\pm$ 0.005	&	2.0	&	2820 $\pm$ 1086	\\
\hline											
	&	Linear	&	\eq{finalUCC} $\rho=1$	&	0.968 $\pm$ 0.068	&	5.1	&	5115 $\pm$ 5475	\\
	&		&	\eq{finalUCC} $\rho=2$	&	0.990 $\pm$ 0.030	&	6.9	&	2882 $\pm$ 3620	\\
	&		&	\eq{exactU}	&	0.997 $\pm$ 0.006	&	6.5	&	4231 $\pm$ 3880	\\
\hline
\multicolumn{6}{l}{$^*$ {\footnotesize Non-parallelism error.}}
\end{tabular}
\end{table*}

\subsubsection{Reduction in the number of parameters by pre-screening of cluster amplitudes}

Classical approximations can provide a criterion to discard excitation operators with small amplitudes, which have a minor contribution to the wavefunction expansion. MP2 amplitudes, for instance, provide an approximation of the amplitudes of double excitation operators, which are responsible for the scaling of the number of parameters in the  UCCSD method as a function of the system size. Given the set of MP2 amplitudes, $\{t_{ij}^{ab(\mathrm{MP2})}\}$, we can discard all the excitation operators such as $|t_{ij}^{ab(\mathrm{MP2})}|<d$, where $d$ is a suitable threshold. \tab{mp2screening} displays the average performance of UCCSD calculations in the H$_4$ systems using a reduced number of parameters for different values of $d$. 

\begin{table*}[h]
\caption{Results of VQE calculations for the H$_4$ systems using prescreening with the MP2 amplitudes. We compared the results obtained for different values of the threshold ($d$) with the calculation including all the parameters. The L-BFGS-B optimization method was used. For $d<10^{-3}$ the results are the same as for $d=10^{-3}$.}\label{tab:mp2screening}
\begin{tabular}{cccccccccccc}
\hline
System & \multicolumn{3}{c}{Number of parameters} && \multicolumn{2}{c}{Max. difference in PES $^{*}$ (kcal/mol)} && \multicolumn{3}{c}{Energy evaluations x $10^3$ } \\
\cline{2-4}\cline{6-7}\cline{9-11}
& $d=10^{-2}$ & $d=10^{-3}$ & All && $d=10^{-2}$ & $d=10^{-3}$ && $d=10^{-2}$ & $d=10^{-3}$ & All\\
\hline
Trapezoidal & 24 & 26 & 52 && $<$7x10-4 & $<$7x10-4 &&  1.17$\pm$0.11 & 1.20$\pm$0.13 & 3.5$\pm$0.6\\
Parallel & 24 & 26 & 52 && 0.07 & $<$7x10-4 && 1.12$\pm$0.44 &  1.17$\pm$0.43 & 2.8$\pm$1.0 \\
Linear & 24 & 26 & 52 && 0.76 & 0.20 && 1.26$\pm$0.43 & 1.37$\pm$0.37 & 4.2$\pm$3.8 \\
\hline
\multicolumn{10}{l}{$^*$ {\footnotesize Maximum difference between the PES calculated with all the parameters and the PES obtained from the pre-screened calculation.}} \\
\hline
\end{tabular}
\end{table*}

For the H$_4$ systems, only 10 out of the 34 double excitation operators have a significant effect on the total energy. The errors in the energy associated to the discarded parameters are in all cases smaller than chemical accuracy. Through the pre-screening process we reduce the circuit depth of the VQE algorithm by reducing the number of parameters involved in the preparation. The smaller number of parameters also facilitates the convergence of the optimization method. For the H$_4$ systems, the number of function evaluations decreases by a factor of 3.

\begin{figure}
\centering
\begin{tabular}{c}  \includegraphics[width=9cm]{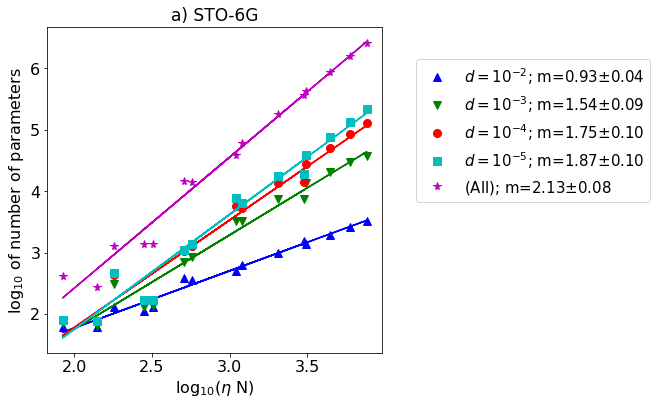} \\ \includegraphics[width=9cm]{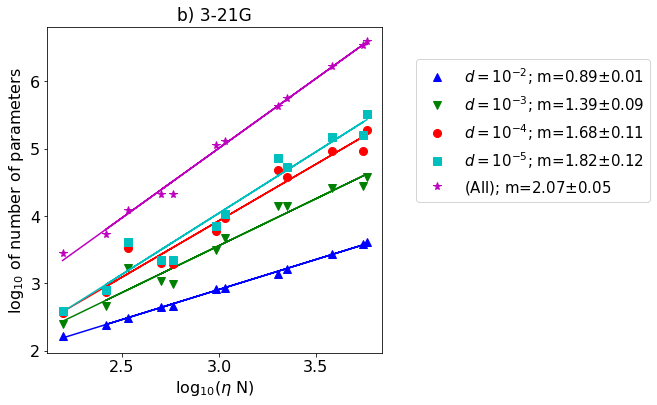} 
\end{tabular}
\caption{Number of parameters in the VQE unitary for different values of the threshold $d$ for some molecules. We employed two different basis sets: a) STO-6G and b) 3-21G. The results are plotted against the product of the number of basis set functions and electrons, $\eta N$. The correlation coefficient (R) and slope (m) of the linear regressions are shown in the legend. The list of molecules include: hydrocarbons (C1-C8), BeH$_2$, Benzene, N$_2$, O$_2$, B$_2$H$_6$, ethanol and water.}\label{fig:mp2screeningmol}
\end{figure}

To gain insights into the effect of the screening in the scaling of the number of parameters for UCC, we applied our reduction strategy to a series of small molecules for different values of the threshold $d$. The results are shown in \fig{mp2screeningmol} as a function of the product of the number of electrons and the number of orbitals of the system, $N\eta$. We observe that the number of parameters, and consequently the depth of the VQE circuit, decreases by almost one order of magnitude using a threshold of $10^{-5}$. In addition, for thresholds above $10^{-5}$, it is also possible to achieve a subquadratic scaling in $N\eta$, compared to the formal quadratic scaling of the full UCCSD ansatz.

\subsubsection{Gradient based optimization}\label{sec:num_grad_analysis}

\begin{figure}
\centering
\includegraphics[width=9cm]{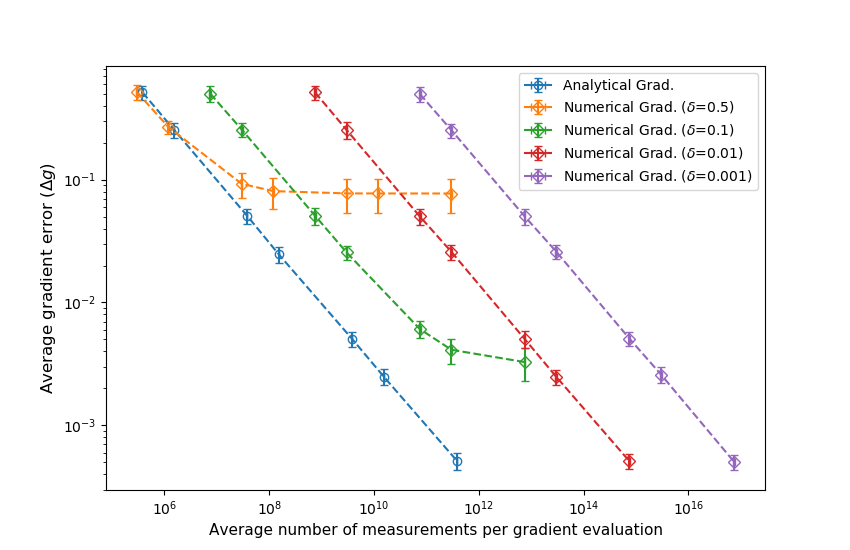}
\caption{Average error of the numerical gradient as a function of the number of measurement employed for the gradient estimation. We compare the analytical gradient and the the numerical gradient for several step sizes. Averages were calculated over 100 random amplitudes drawn uniformly from the interval $[0, 2\pi]$ for the linear $H_4$ system with $r=1.2\AA$. Error bars correspond to one standard deviation.}\label{fig:error_vs_nmeas}
\end{figure}

\begin{figure}
\centering
\includegraphics[width=9cm]{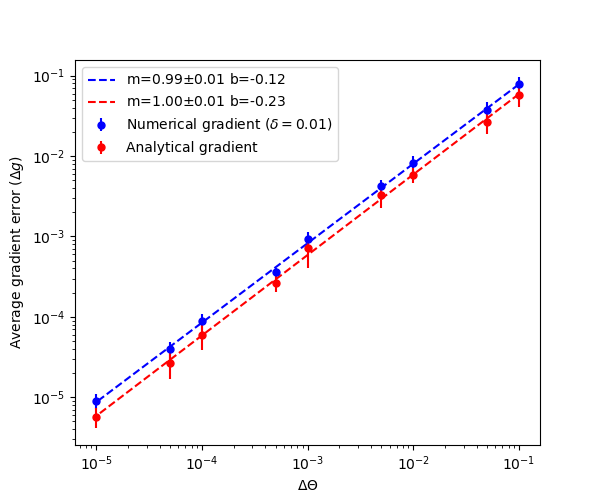}
\caption{Average error of the numerical gradient as a function of the standard deviation of control errors in the quantum circuit ($\Delta\Theta$). $m$ and $b$ correspond to the slope and intercept of the linear regression. Averages were calculated over 100 random amplitudes drawn uniformly from the interval $[0, 2\pi]$ for the linear $H_4$ system with $r=1.2\AA$. The same scaling was observed for other instances of the $H_4$ system.}\label{fig:numvscontrolerrors}
\end{figure}

\begin{table*}[h]
\caption{Average error in the UCC energy (hartrees) and average number of gradient calls employed for the optimization using analytical and numerical gradients (with $\delta=0.05$ and $\delta=0.10$) under the effect of control errors. The error in the energy corresponds to the difference between the optimal energy obtained for 150 VQE runs under control noise and the optimal value for the noiseless optimization with the analytical gradient. All the calculations employed a trotterized ansatz with one trotter step and the same stopping criteria for L-BFGS-B. The UCC amplitudes were initialized with the MP2 amplitudes. The parameter $\Delta\Theta$ was fixed to 0.01. The calculations were performed for instances of the trapezoidal, linear and parallel H$4$ system with the UCC ansatz ($r=1.2\AA$ and $\theta=135.0^o$).}\label{tab:control_e_conv}
\begin{tabular}{cccccccccc}
\hline
& \multicolumn{2}{c}{Trapezoidal} && \multicolumn{2}{c}{Parallel} && \multicolumn{2}{c}{Linear} \\
\cline{2-3}\cline{5-6}\cline{8-9}
& \specialcell{Grad.\\calls} & \specialcell{Energy\\error} && \specialcell{Grad.\\calls} & \specialcell{Energy\\error} && \specialcell{Grad.\\calls} & \specialcell{Energy\\error} \\ 
\hline
Gradient Grad. & 26$\pm$4 & 0.024$\pm$0.008 && 33$\pm$9 & 0.083$\pm$0.086 && 32$\pm$8 & 0.13$\pm$0.08\\
Numerical Grad. ($\delta=0.05$) & 32$\pm$8 & 0.019$\pm$0.006 && 42$\pm$13 & 0.083$\pm$0.095 && 40$\pm$11 & 0.13$\pm$0.08 \\
Numerical Grad. ($\delta=0.1$) & 32$\pm$8 & 0.019$\pm$0.007 && 41$\pm$13 & 0.074$\pm$0.083 && 40$\pm$12 & 0.11$\pm$0.07 \\
\hline
\end{tabular}
\end{table*}

\jrf{Finally, we studied the performance of UCC optimizations with the analytical and numerical gradient approaches proposed in \sec{UCCderivatives}. First, we compared the sampling cost of the analytical and numerical gradient via numerical simulation. We calculated the error in the estimated gradient, $\Delta g$,  as a function of the number of samples employed in the gradient estimation. The gradient error is quantified as the norm of the difference between the estimated gradient, $\tilde{g}$, and the exact gradient, $g$, $\Delta g = ||\tilde{g} - g||_2$. In our numerical simulations, the exact gradient corresponds to the analytical gradient computed to machine precision. To compute the number of measurements, we employed the equality of \eq{nmeas} for the numerical gradient and \eq{nmeas_grad} for the analytical one.

\fig{error_vs_nmeas} illustrates the behavior of the error in the gradient as a function of the number of measurements for a single instance of the $H_4$ system in a linear configuration. Each point in the plot corresponds to an average over 100 gradient estimations for randomly sampled amplitudes.  We compare the numerical gradient with different values of the step size, $\delta$, and the analytical gradient. For $\delta=0.5$ and high error rates, the numerical gradient has a sampling cost comparable to the analytical gradient. However, increasing the sampling cost beyond $10^8$ does not further improve the quality of the numerical approach as the the method reaches its accuracy limit. A similar behavior is observed for the numerical gradient with $\delta=0.1$ for errors below $10^{-3}$. Using smaller step sizes guarantees the same error rate achieved with analytical gradient but with an exceedingly larger sampling cost. Our results confirms the analysis presented in \sec{UCCderivatives}, indicating that the analytical gradient might be order of magnitudes cheaper than numerical gradients in experimental realizations of VQE.

To further understand the relative performance of the analytical and numerical gradients, we numerically simulated the impact of control errors on these methods. Control errors refer to the difference between the formal specification of a variational circuit $U(\vec{p})$, and the actual operation that this specification effects on the quantum computer, $\tilde{U}(\vec{p})$. For simplicity, we will model the control errors as $\tilde{U}(\vec{p}) = U(\vec{p} + \Delta \vec{p})$. In our numerical simulations, $\Delta\vec{p}$ is described as a normal random variable with standard deviation $\Delta\Theta$. 

\fig{numvscontrolerrors} shows the magnitude of $\Delta g$ for the analytical and the numerical gradients as a function of the parameter $\Delta \Theta$. We fixed the value of $\delta$ such as the contribution of the control errors is dominant in the numerical gradient for the ranges of $\Delta\Theta$ explored. Our results show that $\Delta_g$ scales linearly with $\Delta \Theta$, in contrast with the quadratic scaling in $\delta$. In experimental implementations of VQE, $\Delta \Theta$ imposes a practical lower bound to the value of $\delta$ employed in the estimation of the numerical gradient and consequently the contribution of control errors will dominate $\Delta g$. 

Finally, we explore how control errors influence the optimization with numerical and analytical gradients. Assuming that control errors dominate $\Delta g$, we performed 150 runs of the VQE optimization under the influence of control errors, with $\Delta\Theta=0.01$. Our results, summarized in \tab{control_e_conv}, compare the average error in the final energy and average number of gradient calls for carrying out the optimization. We observe that the analytical and the numerical gradients provide accuracies in similar ranges. However, the optimization with analytical gradients requires 20\% less gradient evaluations in average compared to optimization with numerical gradients. These results suggest that the analytical gradient might have a better convergence under the influence of control errors, in addition to a much lower sampling cost.
}

\section{Discussion}\label{sec:discussion}

We have presented a series of strategies for the calculation of molecular energies using the VQE algorithm combined with a UCC ansatz for carrying out the state preparation. The UCC method provides a hierarchy of wavefunction ansatze that can be prepared using quantum circuits with a size that scales polynomially in the number of orbitals and particles of the system. In particular, the approximation up to double cluster operators provides a good compromise between cost and accuracy for applications in quantum  chemistry, with a number of parameters that scales as $O(N^2\eta^2)$. The number of parameters in the approximation determines the size of the circuit and impacts the cost of the classical optimization required for wavefunction optimization. 

Additionally, we have illustrated how efficient classical approximations to the amplitudes of the cluster operators, such as those obtained from perturbation theory, can be used to reduce the cost of the VQE algorithm for chemistry. In particular, we showed that classical amplitudes provide effective initial guesses for the optimization process and serve as a pre-screening mechanism to remove cluster operators that have negligible contribution to the optimal wavefunction. This strategy is an example of a hybrid quantum-classical scheme for quantum simulation, where efficient classical approximations are employed to reduce quantum resources and boost the efficiency of the quantum subroutine. These hybrid schemes are more likely to be the first quantum algorithms to exploit the power of small quantum computers for quantum simulation.

Our numerical analysis also highlights the deficiencies of some derivative-free methods, such as Nelder-Mead and Powell, that have been previously employed in numerical and experimental demonstrations of VQE \cite{Peruzzo.NC.5.4213.2014,Wecker.PRA.92.042303.2015}. These methods performs poorly for a relatively large number of parameters, failing to converge to the correct wavefunctions unless a physically meaningful initial guess is employed. Among the methods tested, COBYLA showed a much better performance.

\jrf{Finally, we introduced an analytical approach to compute the energy gradient for variational circuits and evaluated its performance for the UCC ansatz. This approach allows us to employ gradient based methods to minimize the energy. Our numerical simulations show that our analytical approach provides solutions of the same quality obtained with derivative-free and numerical gradient approaches. In addition, analytical gradients have a much smaller sampling cost than numerical gradients as well as better convergence behavior under the effect of control noise. We point out that our formulation of the analytical gradient can be adapted to other algorithms that employ a quantum-classical hybrid scheme such as the quantum approximate optimization algorithm \cite{farhi.Arxiv.1411.4028.2014} and some methods proposed in the context of quantum machine learning \cite{bang.NJP.16.073017.2014,Wan.arXiv.1612.00104,romero.arXiv.1612.02806}. Future work will be devoted to evaluating the performance of gradient-based and gradient-free optimizations under non-coherent errors and state preparation and measurement (SPAM) errors. 
}

\section*{Acknowledgements}

JR thanks Gian Giacomo Guerreschi for helpful discussions about gradient methods for quantum variational approaches and Libor Veis for helpful comments on the manuscript. JR and PJL acknowledge the Air Force Office of Scientific Research for support under Award: FA9550-12-1-0046. CH acknowledges partial support by the ARC Center of Excellence for Engineered Quantum Systems (Project No. CE110001013). AA-G acknowledges the Army Research Office under Award: W911NF-15-1-0256. The authors thank the Harvard Odyssey cluster facility where the numerical simulations presented in this work were carried out.

\appendix

\section{Commutativity of subterms in excitation operators}\label{app:appendixA}

Assuming real cluster amplitudes, the JW transformation of the single and double cluster operators for UCC can be written as follows:
\begin{align}\label{eq:A1}
t^{a}_{i}(a_a^{\dagger}a_i- h.c.) \equiv \frac{it^{a}_{i}}{2} \bigotimes^{a-1}_{k=i+1}  \sigma^{z}_k (\sigma^{y}_i\sigma^{x}_a-\sigma^{x}_i\sigma^{y}_a)
\end{align}
\begin{align}\label{eq:A2}
t^{ab}_{ij}(a_b^{\dagger} a_a^{\dagger}  a_j a_i - h.c.) \equiv \frac{it^{ab}_{ij}}{8} \bigotimes^{j-1}_{k=i+1} \sigma^{z}_k \bigotimes^{b-1}_{l=a+1} \sigma^{z}_l \notag \\ 
( \sigma^{x}_i \sigma^{x}_j \sigma^{y}_a \sigma^{x}_b +
\sigma^{y}_i \sigma^{x}_j \sigma^{y}_a \sigma^{y}_b \notag \\
+ \sigma^{x}_i \sigma^{y}_j \sigma^{y}_a \sigma^{y}_b 
+ \sigma^{x}_i \sigma^{x}_j \sigma^{x}_a \sigma^{y}_b \notag \\
- \sigma^{y}_i \sigma^{x}_j \sigma^{x}_a \sigma^{x}_b 
- \sigma^{x}_i \sigma^{y}_j \sigma^{x}_a \sigma^{x}_b \notag \\
- \sigma^{y}_i \sigma^{y}_j \sigma^{y}_a \sigma^{x}_b 
- \sigma^{y}_i \sigma^{y}_j \sigma^{x}_a \sigma^{y}_b),
\end{align}
where we assume without lost of generality that $b>a>j>i$. The commutativity among the terms in \eq{A1} and \eq{A2} can be verified by inspection. In general, the JW transformation of an UCC operator of order $n$ will comprise $2^{2n-1}$ terms, composed by the same string of Z operators multiplying the sum of all the possible strings of X and Y operators acting on $2n$ qubits, such as the numbers of X and Y operators are both odd. The commutativity between any of this terms reduces to the commutativity of the strings containing X and Y operators only.

Consider two arbitrary strings of X and Y operators of length $2n$, $P_A=\bigotimes^{2n}_{i=1}\sigma^{a_{i}}_{i}$ and $P_B=\bigotimes^{2n}_{i=1}\sigma^{b_{i}}_{i}$, acting on the same set of qubits. The commutator is given by:
\begin{align}\label{eq:A3}
[P_A,P_B]=\bigotimes^{2n}_{i=1} (\sigma^{a_{i}}_{i} \sigma^{b_{i}}_{i}) - \bigotimes^{2n}_{i=1}(\sigma^{b_{i}}_{i} \sigma^{a_{i}}_{i})
\end{align}
where the product $\sigma^{a_{i}}_{i} \sigma^{b_{i}}_{i}$ can take three values:
\begin{align}\label{eq:A4}
\sigma^{a_{i}}_{i} \sigma^{b_{i}}_{i}=
\left\{
	\begin{array}{ll}
    \bm{1} \quad if \quad a_{i} = b_{i}\\
    i\sigma_z \quad if \quad a_{i} = x \quad b_{i}=y\\
    -i\sigma_z \quad if \quad a_{i} = y \quad b_{i}=x.
	\end{array}
\right.
\end{align}
Applying \eq{A4} to \eq{A3}, we can write:
\begin{align}
&[P_A,P_B]=\nonumber\\ &\left[(-i)^{n^A_y-c_y}(i)^{n^A_x-c_x} - (-i)^{n^B_y-c_y}(i)^{n^B_x-c_x}\right]P,\label{eq:A5}
\end{align}
where $P$ is the string of Pauli matrices obtained from the multiplication and $n^A_x$ and $n^A_y$ are the numbers of X and Y operators in string A, respectively.  $n^B_x$ and  $n^B_y$ are defined analogously. $c_x$ is the number of times $a_i=b_i=x$; $c_y$ is defined accordingly. Rearranging \eq{A5} we obtain:
\begin{align}\label{eq:A6}
[P_A,P_B]= (-1)^{n^A_y-c_y}\left(1-(-1)^{n^B_y-n^A_y}\right)i^{2n-c_x-c_y}P
\end{align}
Now recall that for the UCC operators, $n^A_y$ and $n^B_y$ are both odd and thus $n^B_y-n^A_y$ is even. Consequently, $[P_A,P_B]$ is zero. We conclude that the subterms comprising a single UCC operator commute.

\newpage

\bibliographystyle{apsrev4-1}
\bibliography{biblio,Mendeley}

\end{document}